\documentclass[showpacs,showkeys,amssymb]{revtex4}
\usepackage{graphicx}
\usepackage{dcolumn}
\usepackage{multirow}
\usepackage{bm}

\begin{document}

\title{How fast can an unstable particle decay into two final states and be observed}
%
%
\author{Hua-Xing Chen}
\email{hxchen@buaa.edu.cn}
\affiliation{School of Physics and Nuclear Energy Engineering and International Research Center for Nuclei and Particles in the Cosmos, Beihang University, Beijing 100191, China}
\begin{abstract}
We investigate unstable hadrons (resonances) which mainly decay into two final states in a very short time. We estimate how far at most these two final states can travel away from each other in the half-life of the initial unstable hadron. As examples, this distance is about 2.5 fm for $\rho \rightarrow \pi \pi$, 1.8 fm for $\Delta \rightarrow N \pi$, and 0.6 fm for $f_0(500) \rightarrow \pi \pi$. We calculate this distance for altogether 181 hadrons, among all the 324 ones listed in PDG2012 together with $Z_c(3900)$ and $Z_c(4025)$. We find it is around one femtometer for many hadrons, which is of the same order of magnitude as the hadronic radii. We also estimate this distance for altogether 15 unstable nuclei. We find it is about several femtometers for some of them, which is of the same order of magnitude as the nuclear radii. For example, it is about six femtometers for ${^{12}{\rm O}\rightarrow^{11}{\rm N}p}$.
\end{abstract}
\pacs{11.80.Cr, 13.25.-k, 13.30.-a, 21.10.Tg}
\keywords{resonances, decay process, relativistic dynamics, hadronic radii, nuclear radii}
\maketitle
\pagenumbering{arabic}

\section{Introduction}

There are lots of unstable particles (resonances) observed in particle physics experiments~\cite{Beringer:1900zz}. They have very short lifetimes. For example, the decay width of the $\rho$ meson is around 150 MeV, and its half-life is only about $4.4\times10^{-24}$~s. To calculate their lifetimes (or their decay widths equivalently), theoretical physicists usually use the field theory and the scattering theory based on statistics. However, since their lifetimes are so short that we might find some physics not only in the sense of probability.

In this paper we shall study the decay process of unstable particles. For simplicity, we shall concentrate on those particles which mainly decay into two final states in a very short time. We shall estimate how far at most these two final states can travel away from each other in this short time. We shall not use the field theory nor the scattering theory, but just study their motion using the theory of relativity~\cite{Salam:1962ap,Weinberg:1962hj,Weinberg:1964zza,Weinberg:1965zz,Morgan:1990ct,Baru:2003qq,Hanhart:2011jz,Hyodo:2011qc}. We shall do these calculations at the hadron level so that we can consider the motion of all valence quarks and sea quarks.

Our assumptions are very simple and straightforward: the initial state, an unstable particle $X$, is at rest in the beginning; it has mass $m_X$ and decay width $\Gamma_X$; it decays into two particles $A$ and $B$, having masses $m_A$ and $m_B$, respectively; when $X$ is decaying into $A$ and $B$, the mass difference between the initial and final states, $m_X-m_A-m_B$, is totally and immediately transferred into kinetic energies of $A$ and $B$; this makes they have speeds $v_A$ and $v_B$, in opposite direction. This process can be easily described using the following equations:
\begin{eqnarray}
 {m_A c^2\over\sqrt{1-v_A^2/c^2}} + {m_B c^2\over\sqrt{1-v_B^2/c^2}} &=& m_X c^2\, ,
 \label{eq:motion1}
\\ {m_A\over\sqrt{1-v_A^2/c^2}} v_A + {m_B\over\sqrt{1-v_B^2/c^2}} v_B &=& 0 \, .
\label{eq:motion2}
\end{eqnarray}
Here $c$ is the speed of light. The quantity
\begin{eqnarray}
d_X\equiv{\hbar \over \Gamma_X} (|v_A| + |v_B|) \, ,
\label{def:dx}
\end{eqnarray}
is just the farthest distance that $A$ and $B$ can travel away from each other in the half-life of $X$. We can use the uncertainty principle $\Delta x \Delta p \geq \hbar /2$ to estimate the theoretical uncertainty of $d_X$:
\begin{eqnarray}
\Delta d_X = {\hbar \over 2} {1 \over {m_A\over\sqrt{1-v_A^2/c^2}} |v_A| + {m_B\over\sqrt{1-v_B^2/c^2}} |v_B|} \, ,
\label{def:Deltadx}
\end{eqnarray}

We use the well-established $\rho$ meson and $\Delta$ baryon as examples and show some numbers here. The mass of the $\rho$ meson is measured to be around 770 MeV. It decays into two pions, whose masses are both around 138 MeV. Using Eqs.~(\ref{eq:motion1}) and (\ref{eq:motion2}) we can calculate their speeds: $v_{\pi1} = 0.93 c$ and $v_{\pi2} = - 0.93 c$. The decay width of the $\rho$ meson is around 150 MeV, and so its half-life is around $4.4\times10^{-24}$ s. During this short time, the distance $d_\rho$ that the two pions can travel away from each other at most is around 2.5 fm, and its theoretical error bar from the uncertainty principle is $\Delta d_\rho = 0.14$ fm. The mass of the $\Delta$ baryon is measured to be around 1232 MeV. It decays into one nucleon and one pion, whose masses are around 939 MeV and 138 MeV, respectively. Using Eqs.~(\ref{eq:motion1}) and (\ref{eq:motion2}) we can calculate their speeds: $v_N = 0.24 c$ and $v_\pi = - 0.86 c$. The decay width of the $\Delta$ baryon is around 117 MeV, and so its half-life is around $5.6\times10^{-24}$ s. During this short time, the distance $d_\Delta$ that the nucleon and the pion can travel away from each other at most is around 1.8 fm, and its theoretical error bar from the uncertainty principle is $\Delta d_\Delta = 0.22$ fm.

In this paper we shall calculate the distance $d_X$ for 181 hadrons, among all the 324 ones listed in PDG2012 together with $Z_c(3900)$ and $Z_c(4025)$~\cite{Beringer:1900zz,Ablikim:2013mio,Liu:2013dau,Xiao:2013iha,Ablikim:2013emm}. We shall find that it is smaller for many other hadrons than $d_\rho$ and $d_\Delta$. For example, $d_{\Upsilon(10580)\rightarrow B B}$ is around 1.06$-$1.43 fm, $d_{f_0(500)\rightarrow \pi \pi}$ is around 0.41$-$0.85 fm, and $d_{\Delta(1905)\rightarrow N \rho}$ is around 0.39$-$0.66 fm. As we know, the order of magnitude of the hadronic radii is about one femtometer~\cite{Jamin:2006tj,Oller:2007xd,Amendolia:1986wj,Dally:1982zk,Hou:2001ig,deForcrand:1991kc,Gasser:1983yg}. Particularly, we would like to note that in Ref.~\cite{Albaladejo:2012te} M.~Albaladejo and J.~A.~Oller obtained the quadratic scalar radius of the $\sigma$ meson using the unitary chiral perturbation theory, that is $\langle r^2 \rangle_s^\sigma = (0.19 \pm 0.02) - i(0.06 \pm 0.02)$ fm$^2$. These hadronic radii are sometimes larger than $d_X$, the distance that the two final states can travel away from each other at most in the half-life of $X$. Consequently, if we assume that the initial and final hadrons are all spherical, the two final states might not separate geometrically even after the whole decay process. In this paper we shall discuss this problem and try to find out what kinds of states may not be observed in the experiments due to their small lifetimes.

In this paper we shall also investigate altogether 15 unstable nuclei~\cite{nndc,NRV,wang2012,Audi:2003zz,Atomic2011}, which mainly decay into two final nuclei in a very short time. We shall perform similar calculations and estimate the distance $d_X$ for them. We find it is about several femtometers for some of them, which number is of the same order of magnitude as the nuclear radii. For example, the distance $d_{^{12}{\rm O}\rightarrow ^{11}{\rm N}p}$ is about 5.26$-$6.08 fm, and its theoretical error bar from the uncertainty principle is $\Delta d_{^{12}{\rm O}} = 1.80$ fm.

This paper is organized as follows: in Sec.~\ref{sec:numerical} we shall estimate the distance $d_X$ for some unstable hadrons; the calculations will be done separately for light mesons, heavy mesons, light baryons and heavy baryons in Sec.~\ref{sec:lightmeson}, \ref{sec:heavymeson}, \ref{sec:lightbaryon} and \ref{sec:heavybaryon}, respectively; in Sec.~\ref{sec:unstablenuclei} we shall estimate the distance $d_X$ for some unstable nuclei; Sec.~\ref{sec:summary} is devoted to a short summary and discussions.

\section{Unstable Hadrons}
\label{sec:numerical}

In this section we shall estimate the distance $d_X$ for altogether 181 hadrons, among all the 324 ones listed in PDG2012 together with $Z_c(3900)$ and $Z_c(4025)$~\cite{Beringer:1900zz,Ablikim:2013mio,Liu:2013dau,Xiao:2013iha,Ablikim:2013emm}. We shall separately investigate light mesons, heavy mesons, light baryons and heavy baryons in the following subsections.

\subsection{Light Meson Sector}
\label{sec:lightmeson}

In this subsection we investigate light mesons consisting of $up$, $down$ and $strange$ quarks. We take into account 75 light mesons among all the 101 ones listed in PDG2012~\cite{Beringer:1900zz}. We estimate the distance $d_X$ for them. The results for light unflavored mesons are listed in Tab.~\ref{tab:lightmesons}, and the results for strange mesons are listed in Tab.~\ref{tab:strangemesons}. We also show their decay modes there, where only the spin and parity conservation is taken into account. Other 26 ones are not taken into account, including:
\begin{enumerate}

\item mesons whose decay widths are smaller than 1 MeV: $\pi$, $\eta$, and $K$;

\item mesons which mainly decay into three final states, or only the three-body decay patterns are listed in PDG2012~\cite{Beringer:1900zz}: $\omega(782)$, $\eta^\prime(958)$, $\pi_2(2100)$ and $K(1630)$;

\item mesons whose masses are below the threshold of the (dominant) final states: $f_0(1370)\rightarrow \rho \rho$~\cite{Barberis:1999wn} and $X(1835)\rightarrow p \bar p$~\cite{Abe:2002tw};

\item mesons whose masses and decay patterns are not well known: $f_2(1430)$, $\eta(1760)$, $\rho(1900)$, $\rho_3(1990)$, $f_0(2100)$, $\rho(2150)$, $f_0(2200)$, $\eta(2225)$, $\rho_3(2250)$, $f_4(2300)$, $f_0(2330)$, $\rho_5(2350)$, $K(1460)$, $K_2(1580)$, $K(1830)$, $K_4(2500)$ and $K(3100)$.

\end{enumerate}
Here different isospin partners are counted just once. For example, $\pi^\pm$ and $\pi^0$ are together denoted as $\pi$.

\begin{figure}[hbt]
\begin{center}
\scalebox{0.8}{\includegraphics{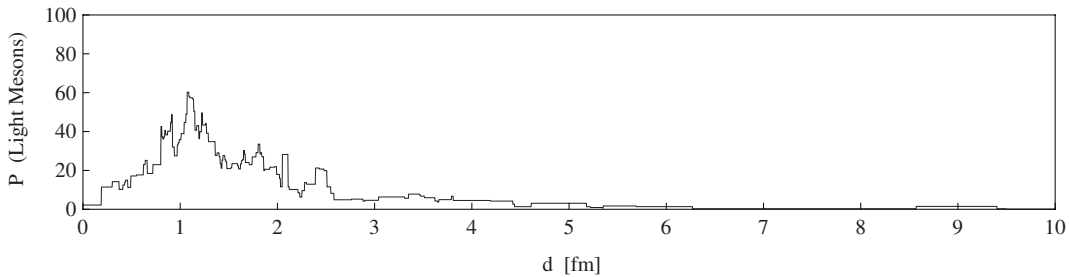}}
\caption{The probability function $P(d)$ for light mesons.}
\label{fig:lightmeson}
\end{center}
\end{figure}

To see the results more clearly, we draw the function $P (d) = \sum_X P_X (d)$ in Fig.~\ref{fig:lightmeson}, where $P_X (d)$ is defined for each light meson $X$:
\begin{eqnarray}
P_X (d) \equiv { H(d- d_X^{lower}) - H(d - d_X^{upper}) \over d_X^{upper} - d_X^{lower} } \, .
\label{def:probability}
\end{eqnarray}
Here $H(x-x_0)$ is the unit step function, and $d_X^{lower}$ and $d_X^{upper}$ are the lower and upper bounds of $d_X$ which have been calculated and listed in Tabs.~\ref{tab:lightmesons} and \ref{tab:strangemesons}. We set $d_X^{lower}=0$ fm and $d_X^{upper}=500$ fm for those having no lower and upper bounds. If the particle $X$ has more than one important (non-negligible) decay partners as explicitly listed in PDG2012~\cite{Beringer:1900zz}, we choose the one having the largest branching ratio and use $^\dagger$ to denote it. This is because it is not very likely that a decay pattern having a larger branching ratio can be fully produced by the final state interaction of a decay pattern having a smaller branching ratio. We can easily verify
\begin{eqnarray}
\int_{0}^{+\infty} P_X (x) dx = 1 \, ,
\end{eqnarray}
and so we call $P(d)$ the probability (density) function.

From its definition Eq.~(\ref{def:probability}) we may guess that: if masses and decay widths are randomly distributed, the probability function $P(d)$ would become larger as $d$ becomes smaller, because $P_X(d)$ is normalized and the denominator $d_X^{upper} - d_X^{lower}$ becomes smaller as $d$ becomes smaller; if we set an upper bound for decay widths, the probability function $P(d)$ would become zero as $d$ goes to zero, due to its definition. Therefore, it should have a maximum, which is related to the (observed) minimum lifetimes of hadrons. From Fig.~\ref{fig:lightmeson} we find its maximum is around one femtometer, which number is of the same order of magnitude as the hadronic radii.

\begin{table}[!hbt]
\begin{center}
\caption{The distance $d_X$ for light unflavored mesons. Some of them have more than one important (non-negligible) decay partners as explicitly listed in PDG2012~\cite{Beringer:1900zz}, and we use $^\dagger$ to denote the one used in the data analysis.}
\begin{tabular}{c | c c | c c c c c | c c | c}
\hline\hline
\multirow{2}{*}{$X$} & Mass & Width & \multirow{2}{*}{$A$} & $M_A$ & \multirow{2}{*}{$B$} & $M_B$ & \multirow{2}{*}{Fraction} & $d_X$ & $\Delta d$ & Decay
\\ & (MeV) & (MeV) & & (MeV) & & (MeV) & & (fm) & (fm) & Modes
\\ \hline\hline
$f_0(500)$ & 400$-$550 & 400$-$700 & $\pi$ & 138 & $\pi$ & 138 & dominant & 0.41$-$0.85 & 0.34 & $S$-wave
\\ \hline
$\rho(770)$ & 761.5$-$775.5 & 146.9$-$154.3 & $\pi$ & 138 & $\pi$ & 138 & 100\% & 2.39$-$2.51 & 0.14 & $P$-wave
\\ \hline
$f_0(980)$ & 990$\pm$20 & 40$-$100 & $\pi$ & 138 & $\pi$ & 138 & dominant & 3.79$-$9.50 & 0.11 & $S$-wave
\\ \hline
$a_0(980)$ & 980$\pm$20 & 50$-$100 & $\pi$ & 138 & $\eta$ & 548 & dominant & 2.76$-$5.69 & 0.16 & $S$-wave
\\ \hline
$\phi(1020)$ & 1019.455$\pm$0.020 & 4.26$\pm$0.04 & $K$ & 496 & $K$ & 496 & 83\% & 21.2$-$21.6 & 0.42 & $P$-wave
\\ \hline
$h_1(1170)$ & 1170$\pm$20 & 360$\pm$40 & $\pi$ & 138 & $\rho$ & 770 & seen & 0.62$-$0.81 & 0.17 & $S$-wave
\\ \hline
$b_1(1235)$ & 1229.5$\pm$3.2 & 142$\pm$9 & $\pi$ & 138 & $\omega$ & 783 & dominant & 1.74$-$1.99 & 0.14 & $S$-wave
\\ \hline
$a_1(1260)$ & 1230$\pm$40 & 250$-$600 & $\pi$ & 138 & $\rho$ & 770 & seen & 0.43$-$1.10 & 0.15 & $S$-wave
\\ \hline
$f_2(1270)$ & 1275.1$\pm$1.2 & 185.1$^{+2.9}_{-2.4}$ & $\pi$ & 138 & $\pi$ & 138 & 85\% & 2.05$-$2.11 & 0.08 & $D$-wave
\\ \hline
$f_1(1285)$ & 1281.9$\pm$0.5 & 24.2$\pm$1.1 & $\pi$ & 138 & $a_0(980)$ & 980 & 36\% & 8.57$-$9.40 & 0.21 & $P$-wave
\\ \hline
$\eta(1295)$ & 1294$\pm$4 & 55$\pm$5 & $\pi$ & 138 & $a_0(980)$ & 980 & seen & 3.66$-$4.44 & 0.20 & $S$-wave
\\ \hline
$\pi(1300)$ & 1300$\pm$100 & 200$-$600 & $\pi$ & 138 & $\rho$ & 770 & seen & 0.44$-$1.47 & 0.15 & $P$-wave
\\ \hline
\multirow{2}{*}{$a_2(1320)$} & \multirow{2}{*}{1322$\pm$7} & \multirow{2}{*}{119$\pm$25} & $\pi$ & 138 & $\rho$ & 770 & -- & 1.96$-$3.02 & 0.12 & $D$-wave
\\ \cline{4-11}
& & & $\pi$ & 138 & $\eta$ & 548 & 15\%~\cite{Shchegelsky:2006et,Forino:1976gq} & 2.28$-$3.51$^\dagger$ & 0.09 & $D$-wave
\\ \hline
$h_1(1380)$ & 1386$\pm$19 & 91$\pm$30 & $K$ & 496 & $K^*(892)$ & 892 & -- & $<$1.04 & $>$0.47 & $S$-wave
\\ \hline
$\pi_1(1400)$ & 1354$\pm$25 & 330$\pm$35 & $\pi$ & 138 & $\eta$ & 548 & seen & 0.90$-$1.13 & 0.09 & $P$-wave
\\ \hline
$\eta(1405)$ & 1408.8$\pm$1.8 & 51.0$\pm$2.9 & $\pi$ & 138 & $a_0(980)$ & 980 & seen & 4.61$-$5.18 & 0.14 & $S$-wave
\\ \hline
$f_1(1420)$ & 1426.4$\pm$0.9 & 54.9$\pm$2.6 & $K$ & 496 & $K^*(892)$ & 892 & dominant & 1.62$-$1.82 & 0.32 & $S$-wave
\\ \hline
$\omega(1420)$ & 1400$-$1450 & 180$-$250 & $\pi$ & 138 & $\rho$ & 770 & dominant & 1.17$-$1.66 & 0.10 & $P$-wave
\\ \hline
$a_0(1450)$ & 1474$\pm$19 & 265$\pm$13 & $\pi$ & 138 & $\eta$ & 548 & seen & 1.22$-$1.36 & 0.08 & $S$-wave
\\ \hline
$\rho(1450)$ & 1465$\pm$25 & 400$\pm$60 & $\pi$ & 138 & $\pi$ & 138 & seen & 0.84$-$1.14 & 0.07 & $P$-wave
\\ \hline
$\eta(1475)$ & 1476$\pm$4 & 85$\pm$9 & $K$ & 496 & $K^*(892)$ & 892 & seen & 1.44$-$1.86 & 0.21 & $P$-wave
\\ \hline
$f_0(1500)$ & 1505$\pm$6 & 109$\pm$7 & $\pi$ & 138 & $\pi$ & 138 & 35\% & 3.35$-$3.81 & 0.07 & $S$-wave
\\ \hline
$f_1(1510)$ & 1518$\pm$5 & 73$\pm$25 & $K$ & 496 & $K^*(892)$ & 892 & seen & 1.64$-$3.47 & 0.17 & $S$-wave
\\ \hline
$f_2^\prime(1525)$ & 1525$\pm$5 & 73$^{+6}_{-5}$ &$K$ & 496 & $K$ & 496 & 89\% & 3.79$-$4.42 & 0.09 & $D$-wave
\\ \hline
$f_2(1565)$ & 1562$\pm$13 & 134$\pm$8 & $\rho$ & 770 & $\rho$ & 770 & seen & 0.30$-$0.66 & 0.59 & $S$-wave
\\ \hline
$\rho(1570)$ & 1570$\pm$36$\pm$62 & 144$\pm$75$\pm$43 & $\pi$ & 138 & $\omega$ & 783 & -- & 1.14$-$12.2 & 0.10 & $P$-wave
\\ \hline
$h_1(1595)$ & 1594$\pm$15$^{+10}_{-60}$ & 384$\pm$60$^{+70}_{-100}$ & $\eta$ & 548 & $\omega$ & 783 & seen & 0.37$-$1.01 & 0.14 & $S$-wave
\\ \hline
$\pi_1(1600)$ & 1662$^{+8}_{-9}$ & 241$\pm$40 & $\pi$ & 138 & $f_1(1285)$ & 1282 & seen & 0.80$-$1.14 & 0.16 & $S$-wave
\\ \hline
$a_1(1640)$ & 1647$\pm$22 & 254$\pm$27 & $\pi$ & 138 & $\rho$ & 770 & seen & 1.13$-$1.41 & 0.08 & $S$-wave
\\ \hline
$f_2(1640)$ & 1639$\pm$6 & 99$^{+60}_{-40}$ & $K$ & 496 & $K$ & 496 & seen & 1.97$-$5.34 & 0.08 & $D$-wave
\\ \hline
$\eta_2(1645)$ & 1617$\pm$5 & 181$\pm$11 & $\pi$ & 138 & $a_2(1320)$ & 1318 & seen & 1.07$-$1.23 & 0.21 & $S$-wave
\\ \hline
$\omega(1650)$ & 1670$\pm$30 & 315$\pm$35 & $\pi$ & 138 & $\rho$ & 770 & seen & 0.91$-$1.15 & 0.08 & $P$-wave
\\ \hline
$\omega_3(1670)$ & 1667$\pm$4 & 168$\pm$10 & $\pi$ & 138 & $\rho$ & 770 & seen & 1.80$-$2.03 & 0.08 & $F$-wave
\\ \hline
$\pi_2(1670)$ & 1672.2$\pm$3.0 & 260$\pm$9 & $\pi$ & 138 & $f_2(1270)$ & 1275 & 56\% & 0.86$-$0.92 & 0.15 & $S$-wave
\\ \hline
$\phi(1680)$ & 1680$\pm$20 & 150$\pm$50 & $K$ & 496 & $K^*(892)$ & 892 & dominant & 1.10$-$2.30 & 0.11 & $P$-wave
\\ \hline
$\rho_3(1690)$ & 1688.8$\pm$2.1 & 161$\pm$10 & $\pi$ & 138 & $\pi$ & 138 & 24\% & 2.28$-$2.58 & 0.06 & $F$-wave
\\ \hline
$\rho(1700)$ & 1720$\pm$20 & 250$\pm$100 & $\pi$ & 138 & $\pi(1300)$ & 1300 & seen & 0.66$-$1.59 & 0.15 & $P$-wave
\\ \hline
$a_2(1700)$ & 1732$\pm$16 & 194$\pm$40 & $\pi$ & 138 & $\rho$ & 770 & -- & 1.38$-$2.12 & 0.07 & $D$-wave
\\ \hline
$f_0(1710)$ & 1720$\pm$6 & 135$\pm$8 & $K$ & 496 & $K$ & 496 & seen & 2.25$-$2.55 & 0.07 & $S$-wave
\\ \hline
$\pi(1800)$ & 1812$\pm$12 & 208$\pm$12 & $\pi$ & 138 & $f_0(500)$ & 500 & seen & 1.65$-$1.86 & 0.06 & $S$-wave
\\ \hline
$f_2(1810)$ & 1815$\pm$12 & 197$\pm$22 & $\pi$ & 138 & $\pi$ & 138 & -- & 1.78$-$2.23 & 0.06 & $D$-wave
\\ \hline
$\phi_3(1850)$ & 1854$\pm$7 & 87$^{+28}_{-23}$ & $K$ & 496 & $K$ & 496 & seen & 2.90$-$5.22 & 0.06 & $F$-wave
\\ \hline
$\eta_2(1870)$ & 1842$\pm$8 & 225$\pm$14 & $\pi$ & 138 & $a_2(1320)$ & 1318 & -- & 1.04$-$1.19 & 0.12 & $S$-wave
\\ \hline
$\pi_2(1880)$ & 1895$\pm$16 & 235$\pm$34 & $\eta$ & 548 & $a_2(1320)$ & 1318 & -- & 0.19$-$0.46 & 0.49 & $S$-wave
\\ \hline
$f_2(1910)$ & 1903$\pm$9 & 196$\pm$31 & $\rho$ & 770 & $\rho$ & 770 & seen & 1.01$-$1.42 & 0.09 & $S$-wave
\\ \hline
$f_2(1950)$ & 1944$\pm$12 & 472$\pm$18 & $\pi$ & 138 & $\pi$ & 138 & seen & 0.80$-$0.86 & 0.05 & $D$-wave
\\ \hline
$f_2(2010)$ & 2011$^{+62}_{-76}$ & 202$^{+67}_{-62}$ & $K$ & 496 & $K$ & 496 & seen & 1.26$-$2.48 & 0.06 & $D$-wave
\\ \hline
$f_0(2020)$ & 1992$\pm$16 & 442$\pm$60 & $\rho$ & 770 & $\rho$ & 770 & seen & 0.49$-$0.66 & 0.08 & $S$-wave
\\ \hline
$a_4(2040)$ & 1996$^{+10}_{-9}$ & 255$^{+28}_{-24}$ & $\pi$ & 138 & $\rho$ & 770 & seen & 1.20$-$1.48 & 0.06 & $G$-wave
\\ \hline
\multirow{2}{*}{$f_4(2050)$} & \multirow{2}{*}{2018$\pm$11} & \multirow{2}{*}{237$\pm$18} & $\omega$ & 783 & $\omega$ & 783 & seen~\cite{Alde:1990qd} & 0.97$-$1.15$^\dagger$ & 0.08 & $D$-wave
\\ \cline{4-11}
& & & $\pi$ & 138 & $\pi$ & 138 & 17\% & 1.53$-$1.79 & 0.05 & $G$-wave
\\ \hline
$f_2(2150)$ & 2157$\pm$12 & 152$\pm$30 & $K$ & 496 & $K$ & 496 & seen & 1.92$-$2.88 & 0.05 & $D$-wave
\\ \hline
$\phi(2170)$ & 2175$\pm$15 & 61$\pm$18 & $f_0(980)$ & 990 & $\phi(1020)$ & 1020 & seen & 1.83$-$3.65 & 0.13 & $S$-wave
\\ \hline
$f_J(2220)$ & 2231.1$\pm$3.5 & 23$^{+8}_{-7}$ & $K$ & 496 & $K$ & 496 & seen & 11.4$-$22.1 & 0.05 & $D$-wave
\\ \hline
$f_2(2300)$ & 2297$\pm$28 & 149$\pm$41 & $K$ & 496 & $K$ & 496 & seen & 1.87$-$3.31 & 0.05 & $D$-wave
\\ \hline
$f_2(2340)$ & 2339$\pm$55 & 319$^{+81}_{-69}$ & $\eta$ & 548 & $\eta$ & 548 & seen & 0.87$-$1.40 & 0.05 & $D$-wave
\\ \hline
$a_6(2450)$ & 2450$\pm$130 & 400$\pm$250 & $K$ & 496 & $K$ & 496 & -- & 0.55$-$2.43 & 0.05 & $I$-wave
\\ \hline
$f_6(2510)$ & 2469$\pm$29 & 283$\pm$40 & $\pi$ & 138 & $\pi$ & 138 & 6\% & 1.22$-$1.62 & 0.04 & $I$-wave
\\ \hline \hline
\end{tabular}
\label{tab:lightmesons}
\end{center}
\end{table}

\begin{table}[!hbt]
\begin{center}
\caption{The distance $d_X$ for strange mesons. Some of them have more than one important (non-negligible) decay partners as explicitly listed in PDG2012~\cite{Beringer:1900zz}, and we use $^\dagger$ to denote the one used in the data analysis.}
\begin{tabular}{c | c c | c c c c c | c c | c}
\hline\hline
\multirow{2}{*}{$X$} & Mass & Width & \multirow{2}{*}{$A$} & $M_A$ & \multirow{2}{*}{$B$} & $M_B$ & \multirow{2}{*}{Fraction} & $d_X$ & $\Delta d$ & Decay
\\ & (MeV) & (MeV) & & (MeV) & & (MeV) & & (fm) & (fm) & Modes
\\ \hline\hline
$K_0^*(800)$ & 682$\pm$29 & 547$\pm$24 & $\pi$ & 138 & $K$ & 496 & -- & 0.19$-$0.37 & 0.76 & $S$-wave
\\ \hline
$K^*(892)$ & 891.4$-$896.4 & 44.4$-$51.7 & $\pi$ & 138 & $K$ & 496 & 100\% & 5.35$-$6.27 & 0.17 & $P$-wave
\\ \hline
\multirow{2}{*}{$K_1(1270)$} & \multirow{2}{*}{1272$\pm$7} & \multirow{2}{*}{90$\pm$20} & $\rho$ & 770 & $K$ & 496 & 42\%~\cite{Daum:1981hb} & $<$0.82$^\dagger$ & $>$0.56 & $S$-wave
\\ \cline{4-11}
& & & $\pi$ & 138 & $K^*(892)$ & 892 & 16\% & 2.19$-$3.49 & 0.17 & $S$-wave
\\ \hline
$K_1(1400)$ & 1403$\pm$7 & 174$\pm$13 & $\pi$ & 138 & $K^*(892)$ & 892 & 94\% & 1.43$-$1.67 & 0.13 & $S$-wave
\\ \hline
$K^*(1410)$ & 1414$\pm$15 & 232$\pm$21 & $\pi$ & 138 & $K^*(892)$ & 892 & $>$40\% & 1.06$-$1.29 & 0.12 & $P$-wave
\\ \hline
$K_0^*(1430)$ & 1425$\pm$50 & 270$\pm$80 & $\pi$ & 138 & $K$ & 496 & 93\% & 0.98$-$1.84 & 0.08 & $S$-wave
\\ \hline
\multirow{2}{*}{$K_2^*(1430)$} & \multirow{2}{*}{1424.1$-$1433.7} & \multirow{2}{*}{95.8$-$114} & $\pi$ & 138 & $K$ & 496 & 50\%~\cite{Aston:1987ir,Estabrooks:1977xe} & 3.04$-$3.62$^\dagger$ & 0.08 & $D$-wave
\\ \cline{4-11}
& & & $\pi$ & 138 & $K^*(892)$ & 892 & 25\% & 2.38$-$2.85 & 0.12 & $D$-wave
\\ \hline
$K_1(1650)$ & 1650$\pm$50 & 150$\pm$50 & $\phi$ & 1020 & $K$ & 496 & -- & 0.66$-$1.84 & 0.20 & $S$-wave
\\ \hline
\multirow{3}{*}{$K^*(1680)$} & \multirow{3}{*}{1717$\pm$27} & \multirow{3}{*}{322$\pm$110} & $\pi$ & 138 & $K$ & 496 & 39\%~\cite{Aston:1987ir} & 0.83$-$1.71$^\dagger$ & 0.07 & $P$-wave
\\ \cline{4-11}
& & & $\rho$ & 770 & $K$ & 496 & 31\% & 0.61$-$1.28 & 0.09 & $P$-wave
\\ \cline{4-11}
& & & $\pi$ & 138 & $K^*(892)$ & 892 & 30\% & 0.70$-$1.45 & 0.08 & $P$-wave
\\ \hline
$K_2(1770)$ & 1773$\pm$8 & 186$\pm$14 & $\pi$ & 138 & $K_2^*(1430)$ & 1430 & dominant & 1.07$-$1.27 & 0.18 & $S$-wave
\\ \hline
\multirow{4}{*}{$K_3^*(1780)$} & \multirow{4}{*}{1776$\pm$7} & \multirow{4}{*}{159$\pm$21} & $\rho$ & 770 & $K$ & 496 & 31\%~\cite{Aston:1986jb} & 1.53$-$2.02$^\dagger$ & 0.08 & $F$-wave
\\ \cline{4-11}
& & & $\pi$ & 138 & $K^*(892)$ & 892 & 20\% & 1.72$-$2.25 & 0.08 & $F$-wave
\\ \cline{4-11}
& & & $\pi$ & 138 & $K$ & 496 & 19\% & 2.02$-$2.63 & 0.06 & $F$-wave
\\ \cline{4-11}
& & & $\eta$ & 548 & $K$ & 496 & 30\% & 1.77$-$2.32 & 0.07 & $F$-wave
\\ \hline
$K_2(1820)$ & 1816$\pm$13 & 276$\pm$35 & $\pi$ & 138 & $K_2^*(1430)$ & 1430 & seen & 0.72$-$0.94 & 0.16 & $S$-wave
\\ \hline
$K_0^*(1950)$ & 1945$\pm$10$\pm$20 & 201$\pm$34$\pm$79 & $\pi$ & 138 & $K$ & 496 & 52\% & 1.17$-$4.19 & 0.06 & $S$-wave
\\ \hline
$K_2^*(1980)$ & 1973$\pm$8$\pm$25 & 373$\pm$33$\pm$60 & $\rho$ & 770 & $K$ & 496 & possibly seen & 0.64$-$1.09 & 0.07 & $D$-wave
\\ \hline
$K_4^*(2045)$ & 2045$\pm$9 & 198$\pm$30 & $\pi$ & 138 & $K$ & 496 & 10\% & 1.63$-$2.21 & 0.05 & $G$-wave
\\ \hline
$K_2(2250)$ & 2247$\pm$17 & 180$\pm$30 & $f_0(980)$ & 990 & $K^*(892)$ & 892 & -- & 1.01$-$1.46 & 0.08 & $D$-wave
\\ \hline
$K_3(2320)$ & 2324$\pm$24 & 150$\pm$30 & $p$ & 938 & $\bar \Lambda$ & 1116 & -- & 0.99$-$1.60 & 0.10 & $D$-wave
\\ \hline
$K_5^*(2380)$ & 2382$\pm$14$\pm$19 & 178$\pm$37$\pm$32 & $\pi$ & 138 & $K$ & 496 & 6\% & 1.52$-$3.46 & 0.04 & $H$-wave
\\ \hline \hline
\end{tabular}
\label{tab:strangemesons}
\end{center}
\end{table}

\subsection{Heavy Meson Sector}
\label{sec:heavymeson}

\begin{figure}[hbt]
\begin{center}
\scalebox{0.8}{\includegraphics{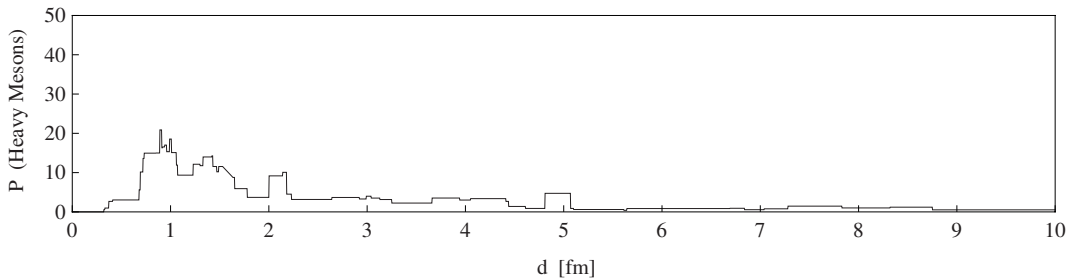}}
\caption{The probability function $P(d)$ for heavy mesons.}
\label{fig:heavymeson}
\end{center}
\end{figure}

In this subsection we investigate heavy mesons containing $charm$ and $bottom$ quarks. We take into account 41 heavy mesons among all the 81 ones listed in PDG2012~\cite{Beringer:1900zz}. We also consider the newly observed mesons $Z_c(3900)$ and $Z_c(4025)$~\cite{Ablikim:2013mio,Liu:2013dau,Xiao:2013iha,Ablikim:2013emm}. We estimate the distance $d_X$ for them. The results for charmed, charmed-strange, bottom, bottom-strange and bottom-charmed mesons are listed in Tab.~\ref{tab:charmedbottommesons}, and the results for charmonium and bottomonium mesons are listed in Tab.~\ref{tab:charmoniumbottomonium}. Other 40 ones are not taken into account, including:
\begin{enumerate}

\item mesons whose decay widths are smaller than 1 MeV: $D^\pm$, $D^0$, $D^{*}(2010)^{\pm}$, $D_s^\pm$, $D_{s1}(2536)^\pm$, $B^\pm$, $B^0$, $B_s^0$, $B_c^\pm$, $J/\psi(1S)$, $\chi_{c1}(1P)$, $h_c(1P)$, $\psi(2S)$, $\Upsilon(1S)$, $\Upsilon(2S)$ and $\Upsilon(3S)$;

\item mesons which mainly decay into three final states, or only the three-body decay patterns are listed in PDG2012~\cite{Beringer:1900zz}: $D^*(2640)^\pm$, $\eta_c(2S)$, $X(4360)$, $X(4660)$ and $\eta_b(1S)$;

\item the meson whose mass is below or too close to the threshold of the (dominant) final states: $X(3872)\rightarrow D^0 \bar D^{*0}$~\cite{Choi:2003ue};

\item mesons whose masses and decay patterns are not well known: $B^*$, $B_1(5721)^0$, $B_s^*$, $B_{s1}(5830)^0$, $B_{sJ}^*(5850)$, $X(4140)$, $\chi_{b0}(1P)$, $\chi_{b1}(1P)$, $h_b(1P)$, $\chi_{b2}(1P)$, $\eta_b(2S)$, $\Upsilon(1D)$, $\chi_{b0}(2P)$, $\chi_{b1}(2P)$, $h_b(2P)$, $\chi_{b2}(2P)$, $\chi_b(3P)$ and $\Upsilon(11020)$.

\end{enumerate}
Here different isospin partners are sometimes counted separately, but the charge-conjugation symmetry is still assumed. For example, $D^{*}(2010)^{\pm}$ and $D^{*}(2007)^{0}$ are counted twice, but $D^{*}(2010)^+$ are $D^{*}(2010)^-$ are together denoted as $D^{*}(2010)^{\pm}$.

We draw the function $P(d) = \sum_X P_X(d)$ in Fig.~\ref{fig:heavymeson}, where $P_X(d)$ is defined for each heavy meson $X$. Again we find it has a maximum around one femtometer. Compared with the case of light mesons, it decreases more slowly as $d$ becomes larger.

\begin{table}[!hbt]
\begin{center}
\caption{The distance $d_X$ for charmed, charmed-strange, bottom, bottom-strange and bottom-charmed mesons. Some of them have more than one important (non-negligible) decay partners as explicitly listed in PDG2012~\cite{Beringer:1900zz}, and we use $^\dagger$ to denote the one used in the data analysis.}
\begin{tabular}{c | c c | c c c c c | c c | c}
\hline\hline
\multirow{2}{*}{$X$} & Mass & Width & \multirow{2}{*}{$A$} & $M_A$ & \multirow{2}{*}{$B$} & $M_B$ & \multirow{2}{*}{Fraction} & $d_X$ & $\Delta d$ & Decay
\\ & (MeV) & (MeV) & & (MeV) & & (MeV) & & (fm) & (fm) & Modes
\\ \hline\hline
\multirow{2}{*}{$D^*(2007)^0$} & \multirow{2}{*}{2006.99$\pm$0.15} & \multirow{2}{*}{$<$2.1} & $\pi^0$ & 135 & $D^0$ & 1865 & 62\%~\cite{Aubert:2005ik} & $>$30.0$^\dagger$ & 1.18 & $P$-wave
\\ \cline{4-11}
& & & $\gamma$ & 0 & $D^0$ & 1865 & 38\% & $>$101 & 0.36 & $P$-wave
\\ \hline
$D_0^*(2400)^0$ & 2318$\pm$29 & 267$\pm$40 & $\pi^\pm$ & 140 & $D^\pm$ & 1870 & seen & 0.72$-$1.01 & 0.14 & $S$-wave
\\ \hline
$D_0^*(2400)^\pm$ & 2403$\pm$14$\pm$35 & 283$\pm$24$\pm$34 & $\pi^\pm$ & 140 & $D^0$ & 1865 & seen & 0.68$-$1.07 & 0.12 & $S$-wave
\\ \hline
$D_1(2420)^0$ & 2421.4$\pm$0.6 & 27.4$\pm$2.5 & $\pi^\pm$ & 140 & $D^{*\pm}$ & 2010 & seen & 7.28$-$8.75 & 0.14 & $S$-wave
\\ \hline
$D_1(2420)^\pm$ & 2423.2$\pm$2.4 & 25$\pm$6 & $\pi^\pm$ & 140 & $D^{*0}$ & 2007 & seen & 7.04$-$11.5 & 0.14 & $S$-wave
\\ \hline
$D_1(2430)^0$ & 2427$\pm$26$\pm$25 & 384$^{+107}_{-75}$$\pm$74 & $\pi^\pm$ & 140 & $D^{*\pm}$ & 2010 & seen & 0.37$-$0.96 & 0.16 & $S$-wave
\\ \hline
$D_2^*(2460)^0$ & 2462.6$\pm$0.6 & 49.0$\pm$1.3 & $\pi^\pm$ & 140 & $D^\pm$ & 1870 & seen & 4.81$-$5.07 & 0.10 & $D$-wave
\\ \hline
$D_2^*(2460)^\pm$ & 2464.3$\pm$1.6 & 37$\pm$6 & $\pi^\pm$ & 140 & $D^0$ & 1865 & seen & 5.64$-$7.83 & 0.10 & $D$-wave
\\ \hline
$D(2550)^0$ & 2539.4$\pm$4.5$\pm$6.8 & 130$\pm$12$\pm$13 & $\pi^\pm$ & 140 & $D^{*\pm}$ & 2010 & seen & 1.49$-$2.23 & 0.11 & $P$-wave
\\ \hline
$D(2600)$ & 2612$\pm$6 & 93$\pm$6$\pm$13 & $\pi$ & 138 & $D^{*}$ & 2009 & seen & 2.14$-$3.25 & 0.10 & --
\\ \hline
$D(2750)$ & 2761$\pm$5 & 63$\pm$6 & $\pi$ & 138 & $D^{*}$ & 2009 & seen & 3.66$-$4.44 & 0.08 & --
\\ \hline \hline
\multirow{2}{*}{$D_{s}^{*\pm}$} & \multirow{2}{*}{2112.3$\pm$0.5} & \multirow{2}{*}{$<$1.9} & $\pi^0$ & 135 & $D_s^\pm$ & 1869 & 6\% & $>$95.3 & 0.26 & $P$-wave
\\ \cline{4-11}
& & & $\gamma$ & 0 & $D_s^\pm$ & 1869 & 94\%~\cite{Aubert:2005ik} & $>$117$^\dagger$ & 0.22 & $P$-wave
\\ \hline
$D_{s0}^*(2317)^\pm$ & 2317.8$\pm$0.6 & $<$3.8 & $\pi^0$ & 135 & $D_s^\pm$ & 1869 & seen & $>$59.6 & 0.13 & $S$-wave
\\ \hline
$D_{s1}(2460)^\pm$ & 2459.6$\pm$0.6 & $<$3.5 & $\pi^0$ & 135 & $D_s^{*\pm}$ & 2112 & 48\% & $>$59.2 & 0.17 & $S$-wave
\\ \hline
$D_{s2}^*(2573)^\pm$ & 2571.9$\pm$0.8 & 17$\pm$4 & $K^\pm$ & 494 & $D^0$ & 1865 & seen & 8.32$-$13.5 & 0.11 & $D$-wave
\\ \hline
\multirow{2}{*}{$D_{s1}^*(2700)^\pm$} & \multirow{2}{*}{2709$\pm$4} & \multirow{2}{*}{117$\pm$13} & $K$ & 496 & $D^*$ & 2009 & -- & 1.30$-$1.65$^\dagger$ & 0.12 & $P$-wave
\\ \cline{4-11}
& & & $K$ & 496 & $D$ & 1867 & -- & 1.59$-$2.00 & 0.09 & $P$-wave
\\ \hline
\multirow{2}{*}{$D_{sJ}^*(2860)^\pm$} & \multirow{2}{*}{2863.2$^{+4.0}_{-2.6}$} & \multirow{2}{*}{58$\pm$11} & $K$ & 496 & $D^*$ & 2009 & -- & 2.99$-$4.41$^\dagger$ & 0.08 & --
\\ \cline{4-11}
& & & $K$ & 496 & $D$ & 1867 & -- & 3.36$-$4.95 & 0.07 & --
\\ \hline
$D_{sJ}(3040)$ & 3044$\pm$8$^{+30}_{-5}$ & 239$\pm$35$^{+46}_{-42}$ & $K$ & 496 & $D^*$ & 2009 & -- & 0.73$-$1.47 & 0.07 & --
\\ \hline \hline
$B_J^*(5732)$ & 5698$\pm$8 & 128$\pm$18 & $\pi$ & 138 & $B^*$ & 5325 & 85\% & 1.33$-$1.78 & 0.15 & --
\\ \hline
\multirow{2}{*}{$B_2^*(5747)^0$} & \multirow{2}{*}{5743$\pm$5} & \multirow{2}{*}{22.7$^{+3.8}_{-3.2}$$^{+3.2}_{-10.2}$} & $\pi^\pm$ & 140 & $B^{*\pm}$ & 5325 & dominant~\cite{Abazov:2007vq,Aaltonen:2008aa} & 6.69$-$21.5$^\dagger$ & 0.13 & $D$-wave
\\ \cline{4-11}
& & & $\pi^\pm$ & 140 & $B^\pm$ & 5279 & dominant & 6.83$-$21.9 & 0.12 & $D$-wave
\\ \hline
$B_{s2}^*(5840)^0$ & 5839.96$\pm$0.20 & 1.56$\pm$0.13$\pm$0.47 & $K^\pm$ & 494 & $B^\pm$ & 5279 & dominant & 46.0$-$104 & 0.20 & $D$-wave
\\ \hline \hline
\end{tabular}
\label{tab:charmedbottommesons}
\end{center}
\end{table}

\begin{table}[!hbt]
\begin{center}
\caption{The distance $d_X$ for charmonium and bottomonium mesons. Some of them have more than one important (non-negligible) decay partners as explicitly listed in PDG2012~\cite{Beringer:1900zz}, and we use $^\dagger$ to denote the one used in the data analysis.}
\begin{tabular}{c | c c | c c c c c | c c | c}
\hline\hline
\multirow{2}{*}{$X$} & Mass & Width & \multirow{2}{*}{$A$} & $M_A$ & \multirow{2}{*}{$B$} & $M_B$ & \multirow{2}{*}{Fraction} & $d_X$ & $\Delta d$ & Decay
\\ & (MeV) & (MeV) & & (MeV) & & (MeV) & & (fm) & (fm) & Modes
\\ \hline\hline
$\eta_c(1S)$ & 2983.7$\pm$0.7 & 32.0$\pm$0.9 & $\rho$ & 770 & $\rho$ & 770 & 1.8\% & 10.3$-$10.9 & 0.04 & $P$-wave
\\ \hline
$\chi_{c0}(1P)$ & 3414.75$\pm$0.31 & 10.3$\pm$0.6 & $\gamma$ & 0 & $J/\psi$ & 3097 & 1.3\% & 19.9$-$22.3 & 0.16 & $S$-wave
\\ \hline
$\chi_{c2}(1P)$ & 3556.20$\pm$0.09 & 1.97$\pm$0.11 & $\gamma$ & 0 & $J/\psi$ & 3097 & 20\% & 108$-$121 & 0.12 & $S$-wave
\\ \hline
$\psi(3770)$ & 3773.15$\pm$0.33 & 27.2$\pm$1.0 & $D$ & 1867 & $D$ & 1867 & 93\% & 2.00$-$2.18 & 0.18 & $P$-wave
\\ \hline
\multirow{2}{*}{$X(3915)$} & \multirow{2}{*}{3918.4$\pm$1.9} & \multirow{2}{*}{20$\pm$5} & $D^0$ & 1865 & $D^{*0}$ & 2007 & -- & 2.38$-$4.12 & 0.17 & $D$-wave
\\ \cline{4-11}
& & & $\omega$ & 783 & $J/\psi$ & 3097 & seen~\cite{Abe:2004zs} & 2.64$-$4.61$^\dagger$ & 0.23 & $S$-wave
\\ \hline
$\chi_{c2}(2P)$ & 3927.2$\pm$2.6 & 24$\pm$6 & $D$ & 1867 & $D$ & 1867 & seen & 4.05$-$6.84 & 0.08 & $D$-wave
\\ \hline
$X(3940)$ & 3942$^{+7}_{-6}\pm$6 & 37$^{+26}_{-15}\pm$8 & $D$ & 1867 & $D^*$ & 2009 & seen & 0.92$-$5.61 & 0.15 & --
\\ \hline
$\psi(4040)$ & 4039$\pm$1 & 80$\pm$10 & $D$ & 1867 & $D^*$ & 2009 & seen & 1.23$-$1.59 & 0.09 & $P$-wave
\\ \hline
$X(4050)^\pm$ & 4051$\pm$14$^{+20}_{-41}$ & 82$^{+21}_{-17}$$^{+47}_{-22}$ & $\pi^\pm$ & 140 & $\chi_{c1}(1P)$ & 3511 & seen & 1.42$-$5.10 & 0.11 & --
\\ \hline
$\psi(4160)$ & 4153$\pm$3 & 103$\pm$8 & $D^*$ & 2009 & $D^*$ & 2009 & seen & 0.89$-$1.06 & 0.10 & $P$-wave
\\ \hline
$X(4160)$ & 4156$^{+25}_{-20}\pm$15 & 139$^{+111}_{-61}\pm$21 & $D^*$ & 2009 & $D^*$ & 2009 & seen & 0.32$-$2.00 & 0.11 & --
\\ \hline
$X(4250)^\pm$ & 4248$^{+44}_{-29}$$^{+180}_{-35}$ & 177$^{+54}_{-39}$$^{+316}_{-61}$ & $\pi^\pm$ & 140 & $\chi_{c1}(1P)$ & 3511 & seen & 0.41$-$3.13 & 0.08 & --
\\ \hline
$X(4260)$ & 4250$\pm$9 & 108$\pm$12 & $f_0(980)$ & 990 & $J/\psi$ & 3097 & seen & 0.99$-$1.30 & 0.10 & $S$-wave
\\ \hline
$X(4350)$ & 4350.6$^{+4.6}_{-5.1}\pm$0.7 & 13$^{+18}_{-9}\pm$4 & $\phi$ & 1020 & $J/\psi$ & 3097 & seen & $>$3.99 & 0.08 & --
\\ \hline
$\psi(4415)$ & 4421$\pm$4 & 62$\pm$20 & $D^*$ & 2009 & $D^*$ & 2009 & not seen & 2.00$-$3.94 & 0.05 & $P$-wave
\\ \hline
$X(4430)^\pm$ & 4443$^{+15}_{-12}$$^{+19}_{-13}$ & 107$^{+86}_{-43}$$^{+74}_{-56}$ & $\pi^\pm$ & 140 & $\psi(2S)$ & 3686 & seen & 0.85$-$28.9 & 0.08 & --
\\ \hline \hline
$\Upsilon(10580)$ & 10579.4$\pm$1.2 & 20.5$\pm$2.5 & $B$ & 5279 & $B$ & 5279 & $>$96\% & 1.06$-$1.43 & 0.15 & $P$-wave
\\ \hline
\multirow{5}{*}{$X(10610)^\pm$} & \multirow{5}{*}{10607.2$\pm$2.0} & \multirow{5}{*}{18.4$\pm$2.4} & $\pi^\pm$ & 140 & $\Upsilon(1S)$ & 9460 & seen~\cite{Belle:2011aa} & 10.5$-$13.6$^\dagger$ & 0.05 & $S$-wave
\\ \cline{4-11}
& & & $\pi^\pm$ & 140 & $\Upsilon(2S)$ & 10023 & seen & 9.72$-$12.7 & 0.09 & $S$-wave
\\ \cline{4-11}
& & & $\pi^\pm$ & 140 & $\Upsilon(3S)$ & 10355 & seen & 8.03$-$10.5 & 0.24 & $S$-wave
\\ \cline{4-11}
& & & $\pi^\pm$ & 140 & $h_b(1P)$ & 9899 & seen & 9.93$-$12.9 & 0.07 & $P$-wave
\\ \cline{4-11}
& & & $\pi^\pm$ & 140 & $h_b(2P)$ & 10260 & seen & 8.94$-$11.7 & 0.16 & $P$-wave
\\ \hline
\multirow{5}{*}{$X(10650)^\pm$} & \multirow{5}{*}{10652.2$\pm$1.5} & \multirow{5}{*}{11.5$\pm$2.2} & $\pi^\pm$ & 140 & $\Upsilon(1S)$ & 9460 & seen~\cite{Belle:2011aa} & 16.0$-$23.6$^\dagger$ & 0.04 & $S$-wave
\\ \cline{4-11}
& & & $\pi^\pm$ & 140 & $\Upsilon(2S)$ & 10023 & seen & 14.9$-$21.9 & 0.08 & $S$-wave
\\ \cline{4-11}
& & & $\pi^\pm$ & 140 & $\Upsilon(3S)$ & 10355 & seen & 13.0$-$19.2 & 0.19 & $S$-wave
\\ \cline{4-11}
& & & $\pi^\pm$ & 140 & $h_b(1P)$ & 9899 & seen & 15.2$-$22.4 & 0.07 & $P$-wave
\\ \cline{4-11}
& & & $\pi^\pm$ & 140 & $h_b(2P)$ & 10260 & seen & 13.9$-$20.5 & 0.14 & $P$-wave
\\ \hline
\multirow{2}{*}{$\Upsilon(10860)$} & \multirow{2}{*}{10876$\pm$11} & \multirow{2}{*}{55$\pm$28} & $B^*$ & 5325 & $B^*$ & 5325 & 38\% & 0.94$-$3.04$^\dagger$ & 0.05 & $P$-wave
\\ \cline{4-11}
& & & $B_s^*$ & 5415 & $B_s^*$ & 5415 & 18\% & 0.38$-$1.50 & 0.11 & $P$-wave
\\ \hline \hline
$Z_c(3900)$ & 3897$\pm$5 & 51$\pm$1 & $D$ & 1867 & $D^*$ & 2009 & -- & 0.69$-$0.91 & 0.28 & --
\\ \hline
$Z_c(4025)$ & 4026.3$\pm$2.6$\pm$3.7 & 24.8$\pm$5.6$\pm$7.7 & $D^*$ & 2009 & $D^*$ & 2009 & -- & 0.33$-$2.92 & 0.78 & --
\\ \hline \hline
\end{tabular}
\label{tab:charmoniumbottomonium}
\end{center}
\end{table}

\subsection{Light Baryon Sector}
\label{sec:lightbaryon}

In this subsection we investigate light baryons consisting of $up$, $down$ and $strange$ quarks. We take into account 53 light baryons among all the 114 ones listed in PDG2012~\cite{Beringer:1900zz}. We estimate the distance $d_X$ for them. The results for light $N$ and $\Delta$ baryons are listed in Tab.~\ref{tab:lightbaryons}, and the results for light $\Lambda$, $\Sigma$, $\Xi$ and $\Omega$ baryons are listed in Tab.~\ref{tab:strangebaryons}. Other 61 ones are not taken into account, including:
\begin{enumerate}

\item baryons whose decay widths are smaller than 1 MeV: $N$, $\Lambda$, $\Sigma$, $\Xi$ and $\Omega$;

\item baryons whose masses and decay patterns are not well known, i.e., all the one-$\star$ and two-$\star$ baryons: $N(1685)$, $N(1860)$, $N(1880)$, $N(1895)$, $N(1990)$, $N(2000)$, $N(2040)$, $N(2060)$, $N(2100)$, $N(2150)$, $N(2300)$, $N(2570)$, $N(2700)$, $N(3000 \mbox{Region})$, $\Delta(1750)$, $\Delta(1900)$, $\Delta(1940)$, $\Delta(2000)$, $\Delta(2150)$, $\Delta(2200)$, $\Delta(2300)$, $\Delta(2350)$, $\Delta(2390)$, $\Delta(2400)$, $\Delta(2750)$, $\Delta(2950)$, $\Delta(3000 \mbox{Region})$, $\Lambda(2000)$, $\Lambda(2020)$, $\Lambda(2325)$, $\Lambda(2585)$, $\Sigma(1480)$, $\Sigma(1560)$, $\Sigma(1580)$, $\Sigma(1620)$, $\Sigma(1620)PE$, $\Sigma(1670)B$, $\Sigma(1690)B$, $\Sigma(1770)$, $\Sigma(1840)$, $\Sigma(1880)$, $\Sigma(2000)$, $\Sigma(2070)$, $\Sigma(2080)$, $\Sigma(2100)$, $\Sigma(2455)B$, $\Sigma(2620)B$, $\Sigma(3000)B$, $\Sigma(3170)B$, $\Xi(1620)$, $\Xi(2120)$, $\Xi(2250)$, $\Xi(2370)$, $\Xi(2500)$, $\Omega(2380)$ and $\Omega(2470)$.

\end{enumerate}
Here different isospin partners are counted just once. For example, $p$ and $n$ are together denoted as $N$.

\begin{figure}[hbt]
\begin{center}
\scalebox{0.8}{\includegraphics{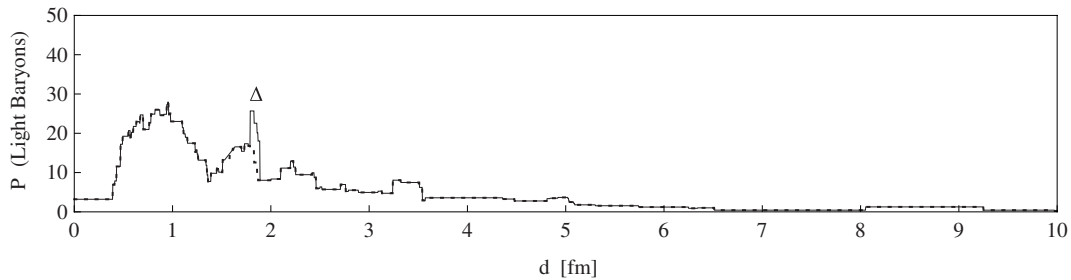}}
\caption{The probability function $P(d)$ for light baryons. The sharp peak around 1.8 fm is due to the $\Delta$ baryon. The dotted curve is obtained by removing this $\Delta$ peak.}
\label{fig:lightbaryon}
\end{center}
\end{figure}

We draw the probability function $P(d) = \sum_X P_X(d)$ in Fig.~\ref{fig:lightbaryon}, where $P_X(d)$ is defined for each light baryon $X$. Again we find it has a maximum around one femtometer. Particulary, the sharp peak around 1.8 fm is due to the $\Delta$ baryon.

\begin{table}[!hbt]
\begin{center}
\caption{The distance $d_X$ for light $N$ and $\Delta$ baryons. Some of them have more than one important (non-negligible) decay partners as explicitly listed in PDG2012~\cite{Beringer:1900zz}, and we use $^\dagger$ to denote the one used in the data analysis.}
\begin{tabular}{c | c c | c c c c c | c c | c}
\hline\hline
\multirow{2}{*}{$X$} & Mass & Width & \multirow{2}{*}{$A$} & $M_A$ & \multirow{2}{*}{$B$} & $M_B$ & \multirow{2}{*}{Fraction} & $d_X$ & $\Delta d$ & Decay
\\ & (MeV) & (MeV) & & (MeV) & & (MeV) & & (fm) & (fm) & Modes
\\ \hline\hline
$N(1440)$ & 1420$-$1470 & 200$-$450 & $\pi$ & 138 & $N$ & 939 & 65\% & 0.58$-$1.34 & 0.13 & $P$-wave
\\ \hline
$N(1520)$ & 1515$-$1525 & 100$-$125 & $\pi$ & 138 & $N$ & 939 & 60\% & 2.20$-$2.76 & 0.11 & $D$-wave
\\ \hline
\multirow{2}{*}{$N(1535)$} & \multirow{2}{*}{1525$-$1545} & \multirow{2}{*}{125$-$175} & $\eta$ & 548 & $N$ & 939 & 42\% & 0.52$-$0.88 & 0.30 & $S$-wave
\\ \cline{4-11}
& & & $\pi$ & 138 & $N$ & 939 & 45\% & 1.58$-$2.23$^\dagger$ & 0.11 & $S$-wave
\\ \hline
$N(1650)$ & 1645$-$1670 & 120$-$180 & $\pi$ & 138 & $N$ & 939 & 70\% & 1.61$-$2.44 & 0.09 & $S$-wave
\\ \hline
\multirow{2}{*}{$N(1675)$} & \multirow{2}{*}{1670$-$1680} & \multirow{2}{*}{130$-$165} & $\pi$ & 138 & $\Delta$ & 1232 & 55\%~\cite{Vrana:1999nt,Anisovich:2011fc} & 1.45$-$1.86$^\dagger$ & 0.14 & $D$-wave
\\ \cline{4-11}
& & & $\pi$ & 138 & $N$ & 939 & 40\% & 1.78$-$2.26 & 0.09 & $D$-wave
\\ \hline
$N(1680)$ & 1680$-$1690 & 120$-$140 & $\pi$ & 138 & $N$ & 939 & 68\% & 2.10$-$2.46 & 0.09 & $F$-wave
\\ \hline
$N(1700)$ & 1650$-$1750 & 100$-$250 & $\pi$ & 138 & $\Delta$ & 1232 & 50\% & 0.95$-$2.52 & 0.14 & $S$-wave
\\ \hline
\multirow{2}{*}{$N(1710)$} & \multirow{2}{*}{1680$-$1740} & \multirow{2}{*}{50$-$250} & $\pi$ & 138 & $\Delta$ & 1232 & 28\% & 0.97$-$5.02$^\dagger$ & 0.13 & $P$-wave
\\ \cline{4-11}
& & & $\eta$ & 548 & $N$ & 939 & 20\% & 0.75$-$4.15 & 0.13 & $P$-wave
\\ \hline
\multirow{2}{*}{$N(1720)$} & \multirow{2}{*}{1700$-$1750} & \multirow{2}{*}{150$-$400} & $\pi$ & 138 & $\Delta$ & 1232 & 75\% & 0.61$-$1.68 & 0.13 & $P$-wave
\\ \cline{4-11}
& & & $\rho$ & 770 & $N$ & 939 & 78\% & $<$0.57$^\dagger$ & $>$0.26 & $P$-wave
\\ \hline
$N(1875)$ & 1820$-$1920 & 160$-$320 & $\pi$ & 138 & $\Delta$ & 1232 & 57\% & 0.82$-$1.70 & 0.10 & $S$-wave
\\ \hline
$N(1900)$ & $\sim$1900 & $\sim$250 & $\omega$ & 783 & $N$ & 939 & 39\% & $\sim$0.67 & 0.12 & $P$-wave
\\ \hline
\multirow{2}{*}{$N(2190)$} & \multirow{2}{*}{2100$-$2200} & \multirow{2}{*}{300$-$700} & $\rho$ & 770 & $N$ & 939 & seen & 0.33$-$0.83 & 0.08 & $D$-wave
\\ \cline{4-11}
& & & $\pi$ & 138 & $N$ & 939 & 15\% & 0.47$-$1.10$^\dagger$ & 0.06 & $G$-wave
\\ \hline
$N(2220)$ & 2200$-$2300 & 350$-$500 & $\pi$ & 138 & $N$ & 939 & 20\% & 0.66$-$0.96 & 0.06 & $H$-wave
\\ \hline
$N(2250)$ & 2200$-$2350 & 230$-$800 & $\pi$ & 138 & $N$ & 939 & 10\% & 0.41$-$1.47 & 0.06 & $G$-wave
\\ \hline
$N(2600)$ & 2550$-$2750 & 500$-$800 & $\pi$ & 138 & $N$ & 939 & 8\% & 0.43$-$0.70 & 0.05 & $I$-wave
\\ \hline \hline
$\Delta(1232)$ & 1230$-$1234 & 114$-$120 & $\pi$ & 138 & $N$ & 939 & 100\% & 1.79$-$1.89 & 0.22 & $P$-wave
\\ \hline
$\Delta(1600)$ & 1500$-$1700 & 220$-$420 & $\pi$ & 138 & $\Delta$ & 1232 & 55\% & 0.47$-$1.11 & 0.24 & $P$-wave
\\ \hline
$\Delta(1620)$ & 1600$-$1660 & 130$-$150 & $\pi$ & 138 & $\Delta$ & 1232 & 45\% & 1.51$-$1.83 & 0.16 & $D$-wave
\\ \hline
\multirow{2}{*}{$\Delta(1700)$} & \multirow{2}{*}{1670$-$1750} & \multirow{2}{*}{200$-$400} & $\pi$ & 138 & $\Delta$ & 1232 & 45\%~\cite{Vrana:1999nt,Anisovich:2011fc} & 0.60$-$1.26$^\dagger$ & 0.14 & $S$-wave
\\ \cline{4-11}
& & & $\rho$ & 770 & $N$ & 939 & 43\% & $<$0.43 & $>$0.26 & $S$-wave
\\ \hline
$\Delta(1905)$ & 1855$-$1910 & 270$-$400 & $\rho$ & 770 & $N$ & 939 & $>$60\% & 0.39$-$0.66 & 0.14 & $P$-wave
\\ \hline
$\Delta(1910)$ & 1860$-$1910 & 220$-$340 & $\pi$ & 138 & $\Delta$ & 1232 & 60\% & 0.78$-$1.23 & 0.10 & $P$-wave
\\ \hline
\multirow{2}{*}{$\Delta(1920)$} & \multirow{2}{*}{1900$-$1970} & \multirow{2}{*}{180$-$300} & $\pi$ & 138 & $N$ & 939 & 13\% & 1.04$-$1.77 & 0.07 & $P$-wave
\\ \cline{4-11}
& & & $\eta$ & 548 & $\Delta$ & 1232 & 15\%~\cite{Anisovich:2011fc} & 0.49$-$0.98$^\dagger$ & 0.16 & $P$-wave
\\ \hline
$\Delta(1930)$ & 1900$-$2000 & 220$-$500 & $\pi$ & 138 & $N$ & 939 & 10\% & 0.63$-$1.45 & 0.07 & $D$-wave
\\ \hline
$\Delta(1950)$ & 1915$-$1950 & 235$-$335 & $\pi$ & 138 & $N$ & 939 & 40\% & 0.94$-$1.35 & 0.07 & $F$-wave
\\ \hline
$\Delta(2420)$ & 2300$-$2500 & 300$-$500 & $\pi$ & 138 & $N$ & 939 & 10\% & 0.67$-$1.15 & 0.05 & $H$-wave
\\ \hline \hline
\end{tabular}
\label{tab:lightbaryons}
\end{center}
\end{table}

\begin{table}[!hbt]
\begin{center}
\caption{The distance $d_X$ for light $\Lambda$, $\Sigma$, $\Xi$ and $\Omega$ baryons. Some of them have more than one important (non-negligible) decay partners as explicitly listed in PDG2012~\cite{Beringer:1900zz}, and we use $^\dagger$ to denote the one used in the data analysis.}
\begin{tabular}{c | c c | c c c c c | c c | c}
\hline\hline
\multirow{2}{*}{$X$} & Mass & Width & \multirow{2}{*}{$A$} & $M_A$ & \multirow{2}{*}{$B$} & $M_B$ & \multirow{2}{*}{Fraction} & $d_X$ & $\Delta d$ & Decay
\\ & (MeV) & (MeV) & & (MeV) & & (MeV) & & (fm) & (fm) & Modes
\\ \hline\hline
$\Lambda(1405)$ & 1405.1$^{+1.3}_{-1.0}$ & 50$\pm$2 & $\pi$ & 138 & $\Sigma$ & 1193 & 100\% & 3.24$-$3.54 & 0.34 & $S$-wave
\\ \hline
\multirow{2}{*}{$\Lambda(1520)$} & \multirow{2}{*}{1519.5$\pm$1.0} & \multirow{2}{*}{15.6$\pm$1.0} & $\pi$ & 138 & $\Sigma$ & 1193 & 42\% & 13.1$-$14.9 & 0.19 & $D$-wave
\\ \cline{4-11}
& & & $K$ & 496 & $N$ & 939 & 45\%~\cite{Corden:1975zz,AlstonGarnjost:1977rs,Gopal:1980ur} & 8.05$-$9.25$^\dagger$ & 0.21 & $D$-wave
\\ \hline
\multirow{2}{*}{$\Lambda(1600)$} & \multirow{2}{*}{1560$-$1700} & \multirow{2}{*}{50$-$250} & $\pi$ & 138 & $\Sigma$ & 1193 & 35\% & 0.91$-$5.04$^\dagger$ & 0.17 & $P$-wave
\\ \cline{4-11}
& & & $K$ & 496 & $N$ & 939 & 23\% & 0.64$-$4.30 & 0.17 & $P$-wave
\\ \hline
\multirow{2}{*}{$\Lambda(1670)$} & \multirow{2}{*}{1660$-$1680} & \multirow{2}{*}{25$-$50} & $\pi$ & 138 & $\Sigma$ & 1193 & 40\% & 4.92$-$9.97$^\dagger$ & 0.13 & $S$-wave
\\ \cline{4-11}
& & & $K$ & 496 & $N$ & 939 & 25\% & 4.04$-$8.35 & 0.12 & $S$-wave
\\ \hline
\multirow{2}{*}{$\Lambda(1690)$} & \multirow{2}{*}{1685$-$1695} & \multirow{2}{*}{50$-$70} & $\pi$ & 138 & $\Sigma$ & 1193 & 30\% & 3.57$-$5.03$^\dagger$ & 0.12 & $D$-wave
\\ \cline{4-11}
& & & $K$ & 496 & $N$ & 939 & 25\% & 3.00$-$4.27 & 0.12 & $D$-wave
\\ \hline
$\Lambda(1800)$ & 1720$-$1850 & 200$-$400 & $K$ & 496 & $N$ & 939 & 33\% & 0.55$-$1.25 & 0.11 & $S$-wave
\\ \hline
\multirow{3}{*}{$\Lambda(1810)$} & \multirow{3}{*}{1750$-$1850} & \multirow{3}{*}{50$-$250} & $K^*(892)$ & 892 & $N$ & 939 & 45\%~\cite{Cameron:1978qi} & $<$1.13$^\dagger$ & $>$0.37 & $P$-wave
\\ \cline{4-11}
& & & $\pi$ & 138 & $\Sigma$ & 1193 & 25\% & 1.04$-$5.42 & 0.11 & $P$-wave
\\ \cline{4-11}
& & & $K$ & 496 & $N$ & 939 & 35\% & 0.91$-$5.01 & 0.10 & $P$-wave
\\ \hline
$\Lambda(1820)$ & 1815$-$1825 & 70$-$90 & $K$ & 496 & $N$ & 939 & 60\%& 2.71$-$3.51 & 0.09 & $F$-wave
\\ \hline
$\Lambda(1830)$ & 1810$-$1830 & 60$-$110 & $\pi$ & 138 & $\Sigma$ & 1193 & 55\% & 2.42$-$4.48 & 0.10 & $D$-wave
\\ \hline
$\Lambda(1890)$ & 1850$-$1910 & 60$-$200 & $K$ & 496 & $N$ & 939 & 28\% & 1.25$-$4.36 & 0.09 & $P$-wave
\\ \hline
$\Lambda(2100)$ & 2090$-$2110 & 100$-$250 & $K$ & 496 & $N$ & 939 & 30\% & 1.15$-$2.89 & 0.07 & $G$-wave
\\ \hline
\multirow{2}{*}{$\Lambda(2110)$} & \multirow{2}{*}{2090$-$2140} & \multirow{2}{*}{150$-$250} & $K^*(892)$ & 892 & $N$ & 939 & 35\%~\cite{Cameron:1978qi} & 0.76$-$1.36$^\dagger$ & 0.10 & $P$-wave
\\ \cline{4-11}
& & & $\pi$ & 138 & $\Sigma$ & 1193 & 25\% & 1.17$-$1.98 & 0.07 & $F$-wave
\\ \hline
\multirow{2}{*}{$\Lambda(2350)$} & \multirow{2}{*}{2340$-$2370} & \multirow{2}{*}{100$-$250} & $\pi$ & 138 & $\Sigma$ & 1193 & 10\% & 1.24$-$3.12 & 0.06 & $H$-wave
\\ \cline{4-11}
& & & $K$ & 496 & $N$ & 939 & 12\%~\cite{deBellefon:1976qr} & 1.24$-$3.13$^\dagger$ & 0.05 & $H$-wave
\\ \hline \hline
$\Sigma(1385)$ & 1382.5$-$1387.7 & 31.0$-$41.5 & $\pi$ & 138 & $\Lambda$ & 1116 & 87\% & 4.81$-$6.51 & 0.24 & $P$-wave
\\ \hline
$\Sigma(1660)$ & 1630$-$1690 & 40$-$200 & $K$ & 496 & $N$ & 939 & 20\% & 0.96$-$5.30 & 0.13 & $P$-wave
\\ \hline
$\Sigma(1670)$ & 1665$-$1685 & 40$-$80 & $\pi$ & 138 & $\Sigma$ & 1193 & 45\% & 3.09$-$6.25 & 0.13 & $D$-wave
\\ \hline
\multirow{2}{*}{$\Sigma(1750)$} & \multirow{2}{*}{1730$-$1800} & \multirow{2}{*}{60$-$160} & $\eta$ & 548 & $\Sigma$ & 1193 & 35\%~\cite{Jones:1974si} & $<$1.77$^\dagger$ & $>$0.23 & $S$-wave
\\ \cline{4-11}
& & & $K$ & 496 & $N$ & 939 & 25\% & 1.40$-$4.01 & 0.11 & $S$-wave
\\ \hline
\multirow{2}{*}{$\Sigma(1775)$} & \multirow{2}{*}{1770$-$1780} & \multirow{2}{*}{105$-$135} & $K$ & 496 & $N$ & 939 & 40\%~\cite{AlstonGarnjost:1977rs,Gopal:1980ur} & 1.73$-$2.25$^\dagger$ & 0.10 & $D$-wave
\\ \cline{4-11}
& & & $\pi$ & 138 & $\Lambda(1520)$ & 1520 & 20\% & 1.38$-$1.81 & 0.26 & $P$-wave
\\ \hline
$\Sigma(1915)$ & 1900$-$1935 & 80$-$160 & $K$ & 496 & $N$ & 939 & 10\% & 1.62$-$3.32 & 0.08 & $F$-wave
\\ \hline
$\Sigma(1940)$ & 1900$-$1950 & 150$-$300 & $K$ & 496 & $N$ & 939 & $<$20\% & 0.87$-$1.79 & 0.08 & $D$-wave
\\ \hline
\multirow{5}{*}{$\Sigma(2030)$} & \multirow{5}{*}{2025$-$2040} & \multirow{5}{*}{150$-$200} & $K$ & 496 & $N$ & 939 & 20\%~\cite{Hemingway:1975uy,Gopal:1980ur} & 1.39$-$1.87$^\dagger$ & 0.07 & $F$-wave
\\ \cline{4-11}
& & & $\pi$ & 138 & $\Lambda$ & 1116 & 20\% & 1.49$-$2.00 & 0.07 & $F$-wave
\\ \cline{4-11}
& & & $\pi$ & 138 & $\Sigma(1385)$ & 1385 & 10\% & 1.31$-$1.75 & 0.09 & $F$-wave
\\ \cline{4-11}
& & & $\pi$ & 138 & $\Lambda(1520)$ & 1520 & 15\% & 1.21$-$1.62 & 0.12 & $D$-wave
\\ \cline{4-11}
& & & $K$ & 496 & $\Delta$ & 1232 & 15\% & 1.06$-$1.44 & 0.10 & $F$-wave
\\ \hline
$\Sigma(2250)$ & 2210$-$2280 & 60$-$150 & $K$ & 496 & $N$ & 939 & $<$10\% & 2.00$-$5.09 & 0.06 & --
\\ \hline \hline
$\Xi(1530)$ & 1531.5$-$1535.6 & 8.0$-$11.6 & $\pi$ & 138 & $\Xi$ & 1318 & 100\% & 14.5$-$21.4 & 0.33 & $P$-wave
\\ \hline
$\Xi(1690)$ & 1690$\pm$10 & $<$30 & $K$ & 496 & $\Lambda$ & 1116 & seen & $>$3.94 & 0.23 & --
\\ \hline
$\Xi(1820)$ & 1823$\pm$5 & 24$^{+15}_{-10}$ & $K$ & 496 & $\Lambda$ & 1116 & large & 4.84$-$13.7 & 0.13 & $D$-wave
\\ \hline
\multirow{2}{*}{$\Xi(1950)$} & \multirow{2}{*}{1950$\pm$15} & \multirow{2}{*}{60$\pm$20} & $K$ & 496 & $\Lambda$ & 1116 & seen~\cite{Biagi:1986vs} & 2.79$-$5.74$^\dagger$ & 0.10 & --
\\ \cline{4-11}
& & & $\pi$ & 138 & $\Xi$ & 1318 & seen & 3.26$-$6.61 & 0.10 & --
\\ \hline
$\Xi(2030)$ & 2025$\pm$5 & 20$^{+15}_{-5}$ & $K$ & 496 & $\Sigma$ & 1193 & 80\% & 6.34$-$14.9 & 0.10 & $D$-wave
\\ \hline \hline
$\Omega(2250)^-$ & 2252$\pm$9 & 55$\pm$18 & $K^-$ & 494 & $\Xi(1530)^0$ & 1530 & seen & 2.50$-$5.08 & 0.12 & --
\\ \hline \hline
\end{tabular}
\label{tab:strangebaryons}
\end{center}
\end{table}

\subsection{Heavy Baryon Sector}
\label{sec:heavybaryon}

In this subsection we investigate heavy baryons containing $charm$ and $bottom$ quarks. We take into account 12 heavy baryons among all the 30 ones listed in PDG2012~\cite{Beringer:1900zz}. We estimate the distance $d_X$ for them. The results are listed in Tab.~\ref{tab:charmedbottombaryons}. Other 18 ones are not taken into account, including:
\begin{enumerate}

\item baryons whose decay widths are smaller than 1 MeV: $\Lambda_c^+$, $\Lambda_c(2625)^+$, $\Xi_c$, $\Omega_c^0$, $\Xi_{cc}$, $\Lambda_b^0$, $\Lambda_b(5912)^0$, $\Lambda_b(5920)^0$, $\Xi_b$ and $\Omega_b^-$;

\item the baryon which mainly decays into three final states, or only the three-body decay patterns are listed in PDG2012~\cite{Beringer:1900zz}: $\Xi_c(2815)$~\cite{Lesiak:2008wz};

\item the baryon whose mass is below or too close to the threshold of the (dominant) final states: $\Lambda_c(2593)^+\rightarrow \Sigma_c(2455) \pi$~\cite{Edwards:1994ar};

\item baryons whose masses and decay patterns are not well known: $\Lambda_c(2765)^+$, $\Xi_c^\prime$, $\Xi_c(2930)$, $\Xi_c(3055)$, $\Xi_c(3123)$ and $\Omega_c(2770)^0$.

\end{enumerate}
Here different isospin partners are counted just once. For example, $\Xi_c^+$ and $\Xi_c^0$ are together denoted as $\Xi_c$.

\begin{figure}[hbt]
\begin{center}
\scalebox{0.8}{\includegraphics{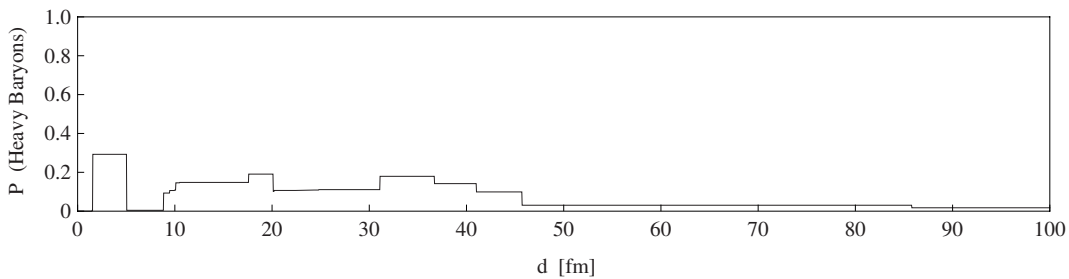}}
\caption{The probability function $P(d)$ for heavy baryons.}
\label{fig:heavybaryon}
\end{center}
\end{figure}

We draw the probability function $P(d) = \sum_X P_X(d)$ in Fig.~\ref{fig:heavybaryon}, where $P_X(d)$ is defined for each heavy baryon $X$. We find it is quite flat probably due to lack of data.

\begin{table}[!hbt]
\begin{center}
\caption{The distance $d_X$ for charmed and bottom baryons. Some of them have more than one important (non-negligible) decay partners as explicitly listed in PDG2012~\cite{Beringer:1900zz}, and we use $^\dagger$ to denote the one used in the data analysis.}
\begin{tabular}{c | c c | c c c c c | c c | c}
\hline\hline
\multirow{2}{*}{$X$} & Mass & Width & \multirow{2}{*}{$A$} & $M_A$ & \multirow{2}{*}{$B$} & $M_B$ & \multirow{2}{*}{Fraction} & $d_X$ & $\Delta d$ & Decay
\\ & (MeV) & (MeV) & & (MeV) & & (MeV) & & (fm) & (fm) & Modes
\\ \hline\hline
\multirow{2}{*}{$\Lambda_c(2880)^+$} & \multirow{2}{*}{2881.53$\pm$0.35} & \multirow{2}{*}{5.8$\pm$1.1} & $\pi^\pm$ & 140 & $\Sigma_c$ & 2455 & seen~\cite{Abe:2006rz} & 31.1$-$45.7$^\dagger$ & 0.13 & $F$-wave
\\ \cline{4-11}
& & & $D^0$ & 1865 & $p$ & 938 & seen & 13.9$-$20.5 & 0.16 & $F$-wave
\\ \hline
\multirow{2}{*}{$\Lambda_c(2940)^+$} & \multirow{2}{*}{2939.3$^{+1.4}_{-1.5}$} & \multirow{2}{*}{17$^{+8}_{-6}$} & $\pi^\pm$ & 140 & $\Sigma_c$ & 2455 & seen~\cite{Abe:2006rz} & 8.84$-$20.1$^\dagger$ & 0.12 & --
\\ \cline{4-11}
& & & $D^0$ & 1865 & $p$ & 938 & seen & 4.94$-$11.3 & 0.12 & --
\\ \hline\hline
$\Sigma_c(2455)$ & 2452.5$-$2454.1 & $<$4.6 & $\pi$ & 138 & $\Lambda_c^+$ & 2287 & 100\% & $>$24.8 & 0.56 & $P$-wave
\\ \hline
$\Sigma_c(2520)$ & 2515.2$-$2519.8 & $<$17 & $\pi$ & 138 & $\Lambda_c^+$ & 2287 & 100\% & $>$9.97 & 0.29 & $P$-wave
\\ \hline
$\Sigma_c(2800)$ & 2787$-$2811 & 1$-$151 & $\pi$ & 138 & $\Lambda_c^+$ & 2287 & seen & 1.49$-$228 & 0.11 & --
\\ \hline\hline
$\Xi_c(2645)$ & 2645.3$-$2646.4 & $<$5.5 & $\pi^\pm$ & 140 & $\Xi_c$ & 2469 & seen & $>$22.9 & 0.48 & $P$-wave
\\ \hline
$\Xi_c(2790)$ & 2785.9$-$2795.1 & $<$15 & $\pi$ & 138 & $\Xi_c^\prime$ & 2577 & seen & $>$10.5 & 0.33 & $S$-wave
\\ \hline
$\Xi_c(2980)$ & 2965.4$-$2974.7 & 13$-$33 & $K$ & 496 & $\Sigma_c$ & 2455 & seen & 1.56$-$5.03 & 0.45 & --
\\ \hline
$\Xi_c(3080)$ & 3076.6$-$3081.3 & 3.4$-$7.8 & $K$ & 496 & $\Sigma_c$ & 2455 & seen & 17.6$-$41.0 & 0.15 & --
\\ \hline\hline
$\Sigma_b$ & 5808.8$-$5817.8 & 1.7$-$14.7 & $\pi$ & 138 & $\Lambda_b^0$ & 5619 & dominant & 9.45$-$85.8 & 0.39 & $P$-wave
\\ \hline
$\Sigma_b^*$ & 5829.6$-$5837.4 & 4.3$-$15.2 & $\pi$ & 138 & $\Lambda_b^0$ & 5619 & dominant & 10.1$-$36.7 & 0.32 & $P$-wave
\\ \hline\hline
$\Xi_b(5945)$ & 5945.5$\pm$0.8$\pm$2.2 & 2.1$\pm$1.7 & $\pi^+$ & 140 & $\Xi_b^-$ & 5791 & seen & 20.2$-$230 & 0.86 & $P$-wave
\\ \hline \hline
\end{tabular}
\label{tab:charmedbottombaryons}
\end{center}
\end{table}

\section{Unstable Nuclei}
\label{sec:unstablenuclei}

\begin{figure}[hbt]
\begin{center}
\scalebox{0.8}{\includegraphics{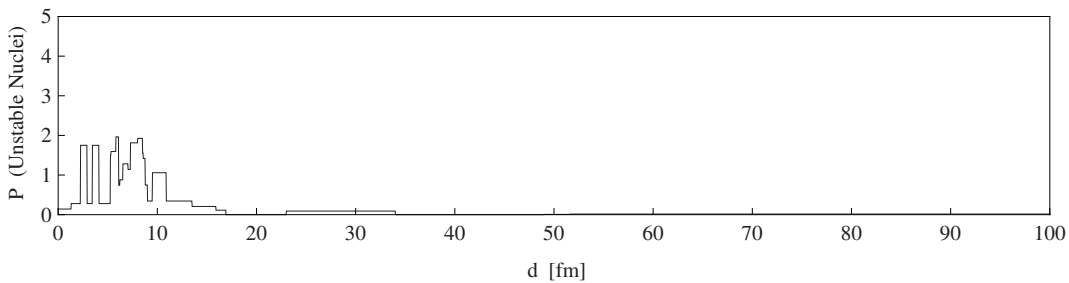}}
\caption{The probability function $P(d)$ for some unstable nuclei.}
\label{fig:unstablenuclei}
\end{center}
\end{figure}

In this section we investigate some unstable nuclei, and estimate the distance $d_X$ for them~\cite{nndc,NRV,wang2012,Audi:2003zz,Atomic2011}. We select altogether 15 unstable nuclei, which have very short lifetimes. All of them decay by emitting one proton or one neutron. The results are listed in Tab.~\ref{tab:unstablenuclei}. We note that we have used atomic masses, which makes little difference compared to the use of nuclear masses. We find that in many cases the distance $d_X$ is about several femtometres, which number is of the same order of magnitude as the nuclear radii.

We draw the probability function $P(d) = \sum_X P_X(d)$ in Fig.~\ref{fig:unstablenuclei}, where $P_X(d)$ is defined for each unstable nucleus $X$ listed in Tab.~\ref{tab:unstablenuclei}. Because there are many other unstable nuclei which are not investigated in this paper, we shall not discuss this figure any more.

\begin{table}[!hbt]
\begin{center}
\caption{The distance $d_X$ for some unstable nuclei. 1 $u$ = $931.494061(21)$ MeV/c$^2$.}
\begin{tabular}{c | c c | c c c c c | c c }
\hline\hline
\multirow{2}{*}{$X$} & Mass & Width & \multirow{2}{*}{$A$} & $M_A$ & \multirow{2}{*}{$B$} & $M_B$ & \multirow{2}{*}{Fraction} & $d_X$ & $\Delta d$
\\ & (MeV) & (MeV) & & (MeV) & & (MeV) & & (fm) & (fm)
\\ \hline\hline
$^4$H~\cite{Gurov:2005sp,TerAkopian} & $4.02781(11)$ & $1.39\pm0.10$ & $n$ & $1.00867$ & $^3$H & $3.01605$ & -- & 3.43$-$4.11 & 0.79
\\ \hline
$^5$He & $5.01222(5)$ & $7.0\pm0.3$ & $n$ & $1.00867$ & $^4$He & $4.00260$ & -- & 9.50$-$10.9 & 1.39
\\ \hline
$^7$He~\cite{Korsheninnikov:2003bz} & $7.02802(2)$ & $29\pm5$ & $n$ & $1.00867$ & $^6$He & $6.01889$ & -- & 23.0$-$34.0 & 1.92
\\ \hline
$^9$He & $9.04395(3)$ & $70\pm40$ & $n$ & $1.00867$ & $^8$He & $8.03392$ & -- & 49.0$-$184 & 1.09
\\ \hline
$^4$Li & $4.02719(23)$ & $0.91\pm0.09$ & $^1$H & $1.00783$ & $^3$He & $3.01603$ & -- & 2.23$-$2.91 & 0.77
\\ \hline
$^5$Li & $5.01254(5)$ & $3.7\pm0.3$ & $^1$H & $1.00783$ & $^4$He & $4.00260$ & -- & 7.29$-$8.78 & 0.92
\\ \hline
$^{10}$Li & $10.03548(2)$ & $20\pm5$ & $n$ & $1.00867$ & $^9$Li & $9.02679$ & -- & $<$7.04 & $>$6.22
\\ \hline
$^{13}$Be & $13.03569(8)$ & $\sim27$ & $n$ & $1.00867$ & $^{12}$Be & $12.02692$ & -- & 5.31$-$15.9 & 8.69
\\ \hline
$^7$B & $7.02992(8)$ & $3.5\pm0.5$ & $^1$H & $1.00783$ & $^6$Be & $6.01973$ & -- & 6.53$-$9.01 & 0.84
\\ \hline
$^{10}$N & $10.04165(43)$ & $2.0\pm1.4$ & $^1$H & $1.00783$ & $^9$C & $9.03104$ & -- & 1.29$-$8.57 & 0.81
\\ \hline
$^{11}$N & $11.02609(5)$ & $5.9\pm2.1$ & $^1$H & $1.00783$ & $^{10}$C & $10.01685$ & -- & 6.21$-$13.5 & 1.06
\\ \hline
$^{12}$O & $12.03441(2)$ & $5.8\pm0.3$ & $^1$H & $1.00783$ & $^{11}$N & $11.02609$ & -- & 5.26$-$6.08 & 1.80
\\ \hline
$^{15}$F & $15.01801(14)$ & $4.1\pm0.6$ & $^1$H & $1.00783$ & $^{14}$O & $14.00860$ & -- & 5.81$-$8.52 & 1.02
\\ \hline
$^{16}$F & $16.01147(1)$ & $110\pm60$ & $^1$H & $1.00783$ & $^{15}$O & $15.00307$ & -- & 51.6$-$179 & 1.63
\\ \hline
$^{18}$Na & $18.02597(5)$ & $13\pm4$ & $^1$H & $1.00783$ & $^{17}$Ne & $17.01767$ & -- & 8.02$-$16.9 & 1.87
\\ \hline \hline
\end{tabular}
\label{tab:unstablenuclei}
\end{center}
\end{table}

\section{Discussions and Summary}
\label{sec:summary}

\begin{figure}[hbt]
\begin{center}
\scalebox{0.8}{\includegraphics{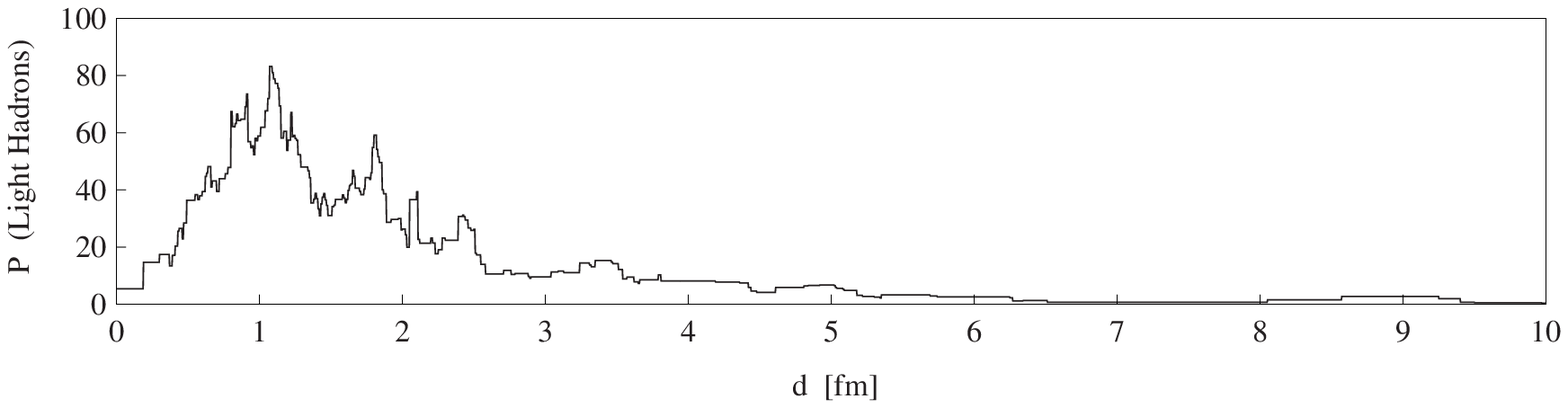}}
\scalebox{0.8}{\includegraphics{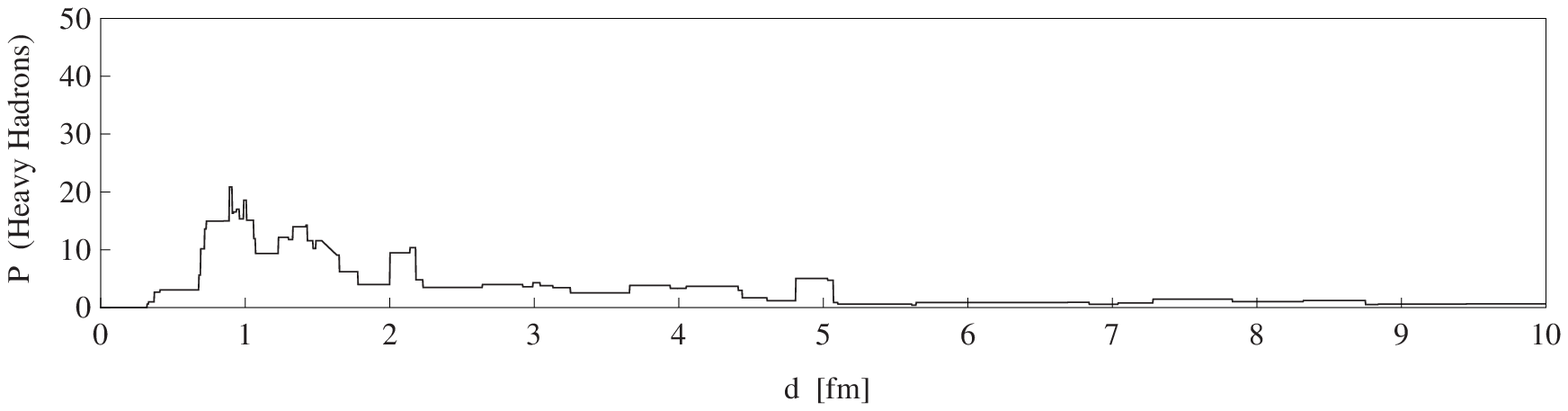}}
\scalebox{0.8}{\includegraphics{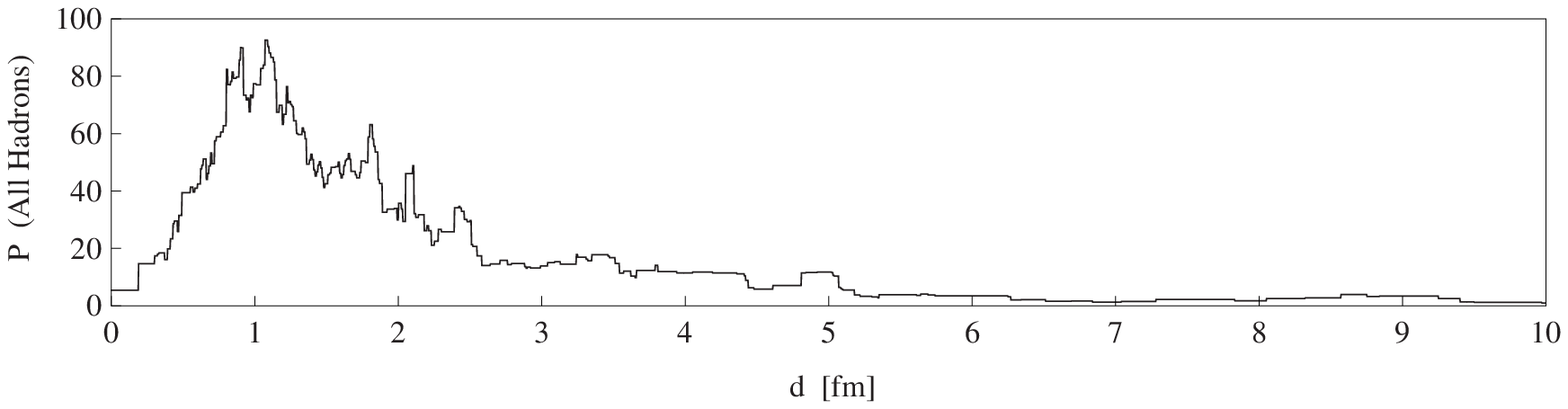}}
\caption{The probability function $P(d)$. The top, middle and bottom figures are for light hadrons, heavy hadrons and all hadrons, respectively.}
\label{fig:totalhadron}
\end{center}
\end{figure}

To summarize this paper, we have investigated unstable hadrons (resonances) which mainly decay into two final states in a very short time. We have taken into account 181 hadrons, among all the 324 ones listed in PDG2012 together with $Z_c(3900)$ and $Z_c(4025)$. We have estimated how far at most the two final states can travel away from each other in the half-life of the initial unstable hadron. Synthesizing all the data obtained in Sec.~\ref{sec:numerical}, we draw the probability function $P(d) = \sum_X P_X(d)$ at the bottom of Figs.~\ref{fig:totalhadron}, where $P_X(d)$ is defined for each unstable hadron $X$ investigated in this paper. We find the distance $d_X$ is around one femtometer for many hadrons, which number is of the same order of magnitude as the hadronic radii. We have also estimated this distance $d_X$ for altogether 15 unstable nuclei. We find that in many cases it is about several femtometres, which number is of the same order of magnitude as the nuclear radii.

\begin{figure}[hbt]
\begin{center}
\scalebox{0.6}{\includegraphics{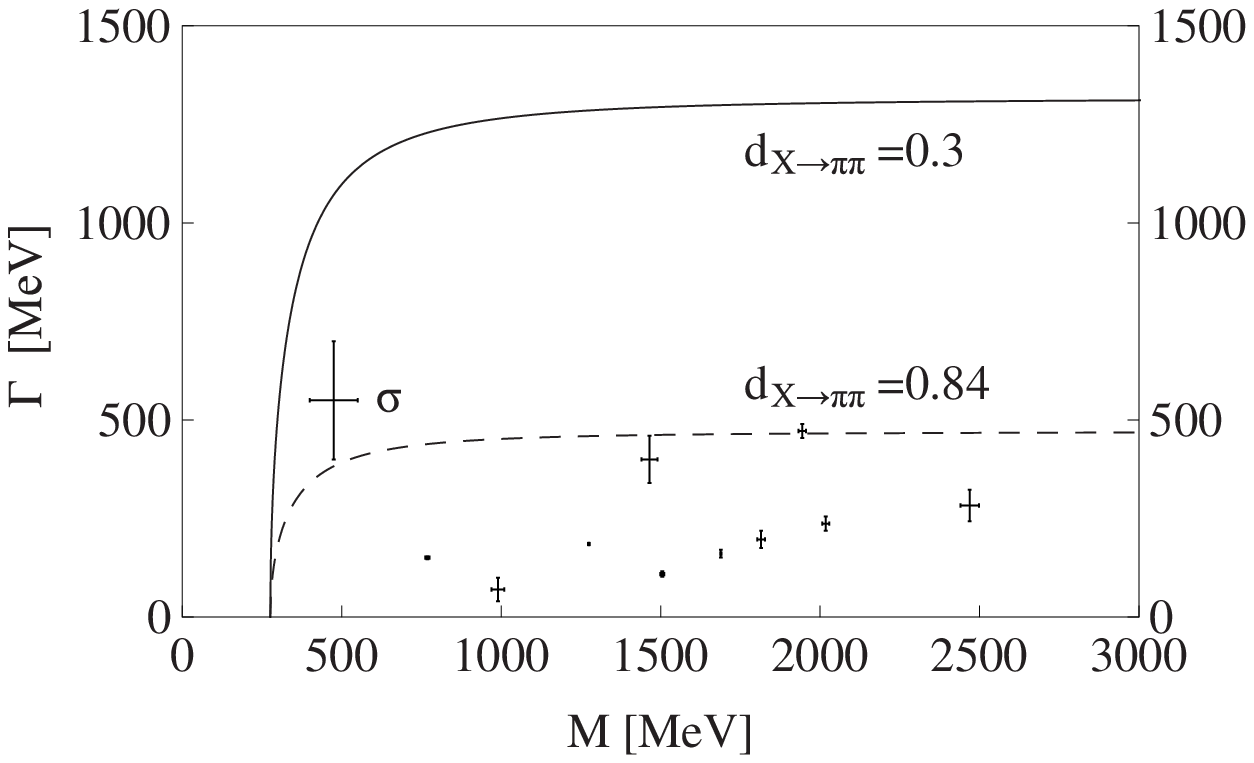}}
\scalebox{0.6}{\includegraphics{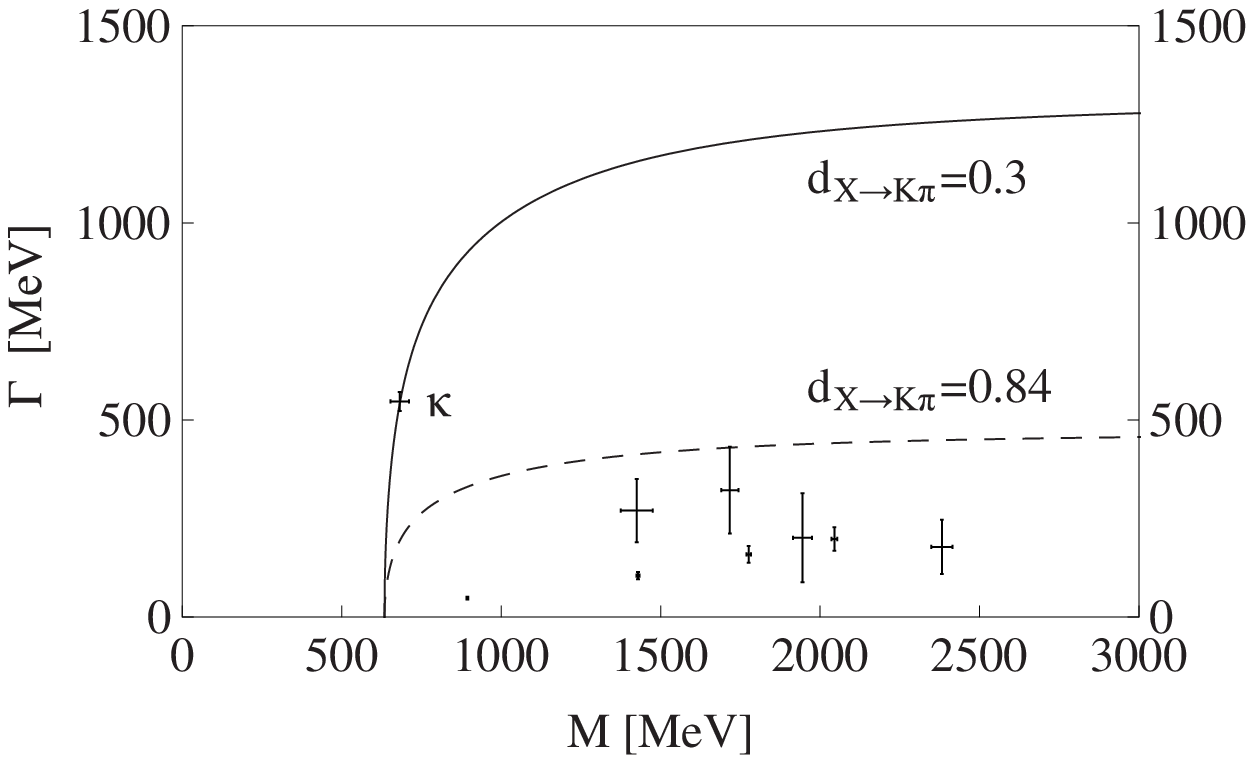}}
\scalebox{0.6}{\includegraphics{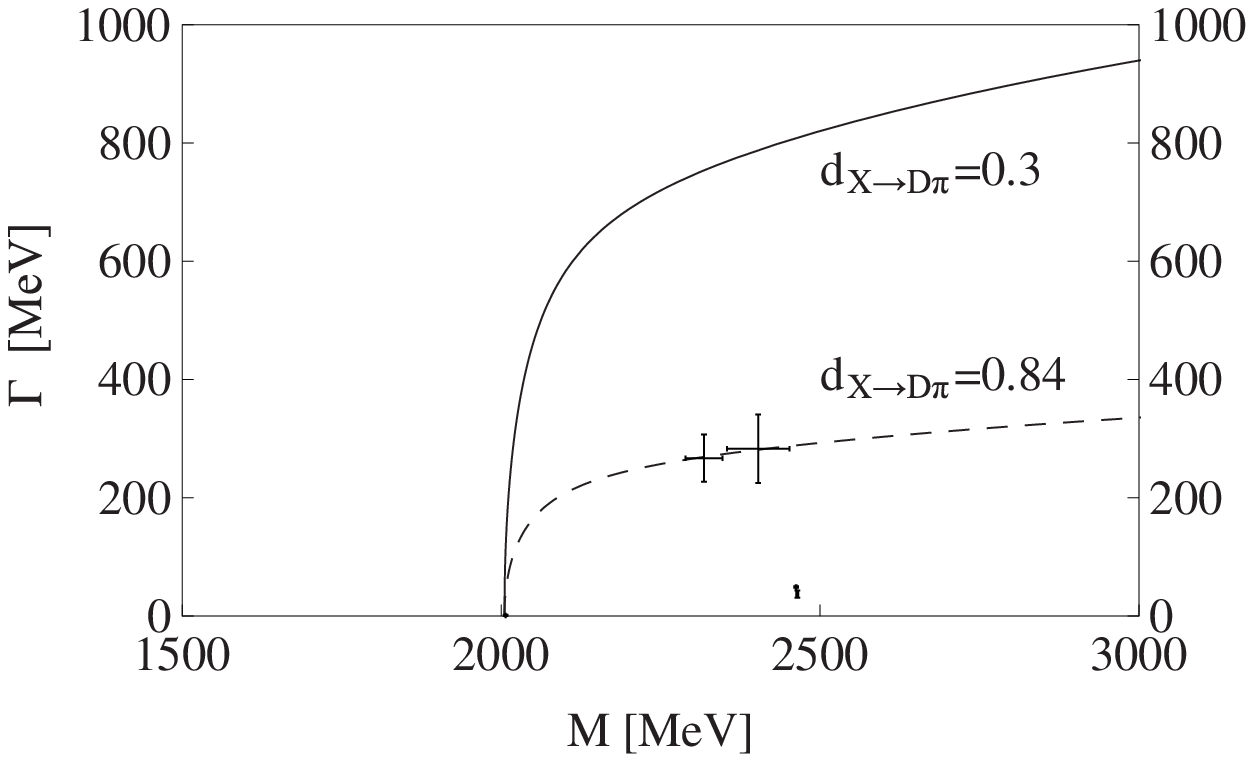}}
\scalebox{0.6}{\includegraphics{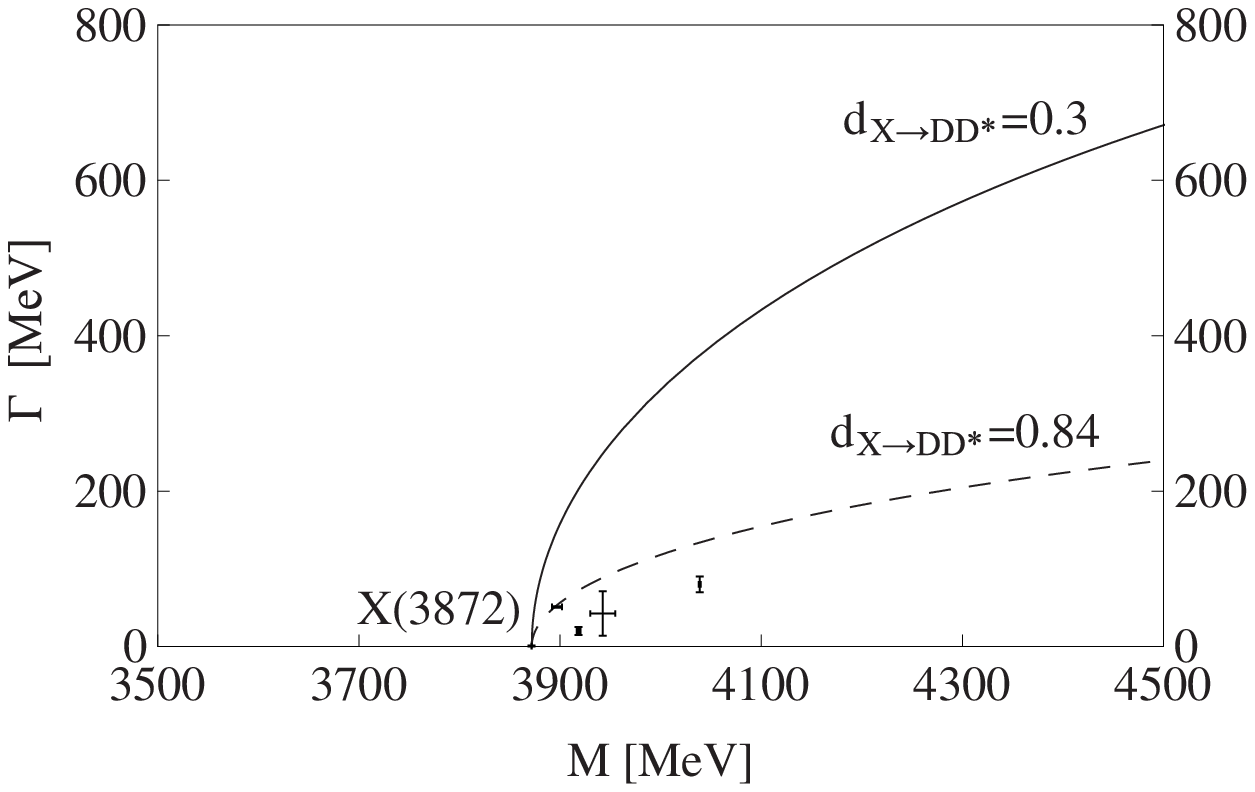}}
\scalebox{0.6}{\includegraphics{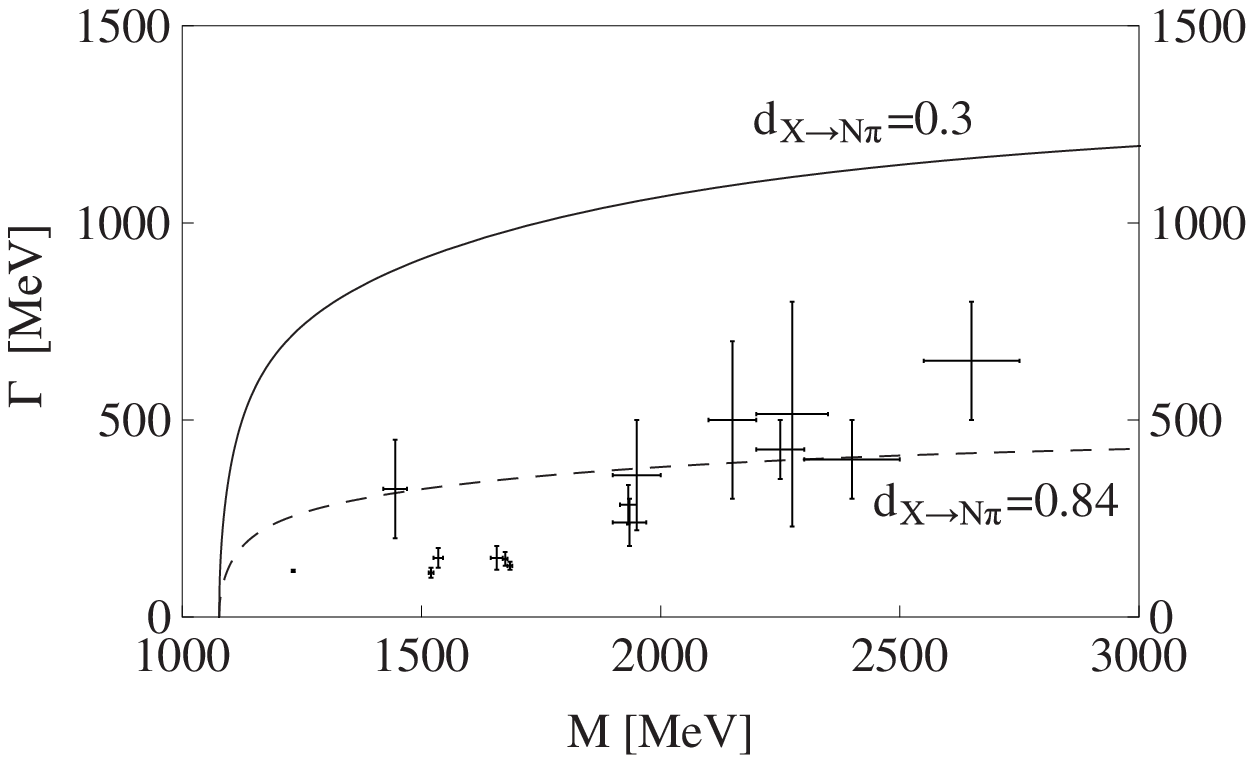}}
\scalebox{0.6}{\includegraphics{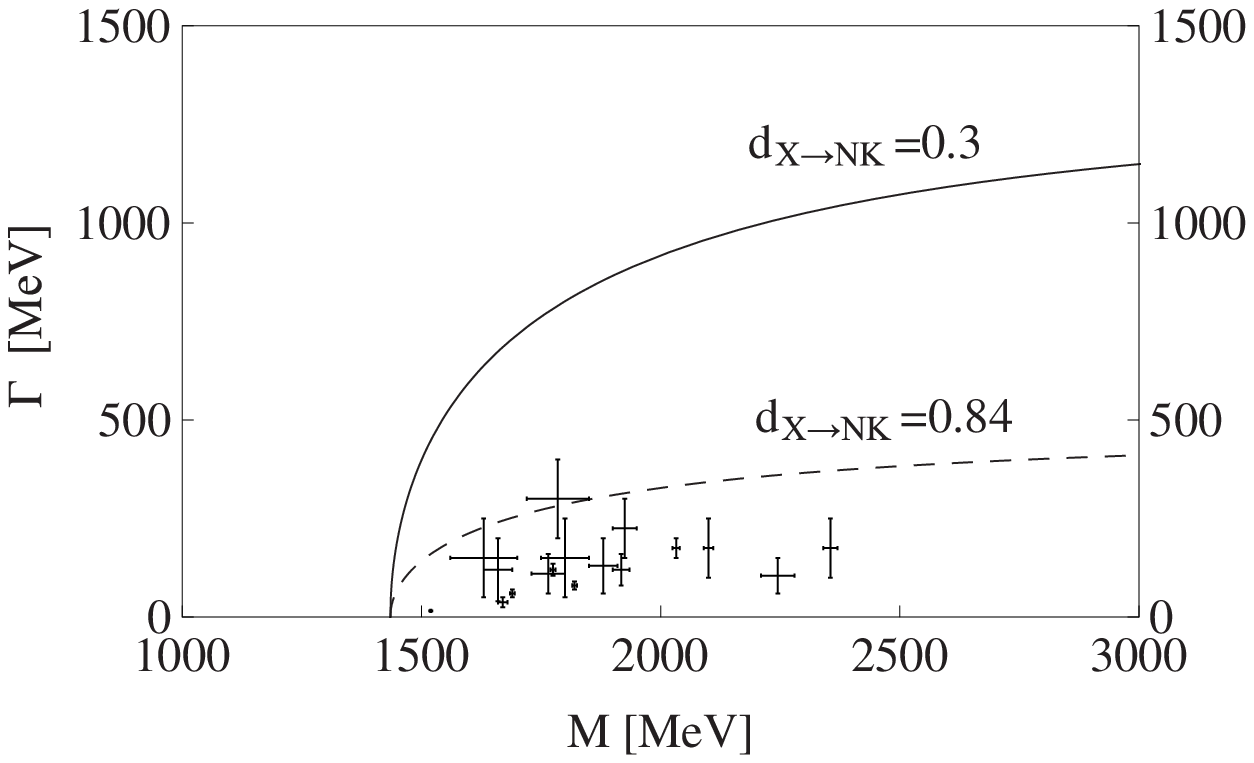}}
\caption{The relation between decay widths and masses. The solid and dashed curves are obtained by fixing $d_X = 0.3$ fm and $d_X = r_p = 0.84$ fm, respectively. The top-left, top-right, middle-left, middle-right, bottom-left and bottom-right figures are obtained by setting the two final states to be $\pi$-$\pi$, $K$-$\pi$, $D$-$\pi$, $D^0$-$D^{*0}$, $N$-$\pi$ and $N$-$K$, respectively. Particularly we show $X(3872)$ in the middle-right figure, where $m_{D^0} = 1864.86$ MeV, $m_{D^{*0}} = 2006.99$ MeV, $m_{X(3872)} = 3871.68 \pm 0.17$ MeV and $\Gamma_{X(3872)} < 1.2$ MeV. It is coincident with the beginnings of the curves $d_{X \rightarrow D D^*} = 0.3$ fm and $d_{X \rightarrow D D^*} = 0.84$ fm.}
\label{fig:limitation}
\end{center}
\end{figure}

\begin{figure}[hbt]
\begin{center}
\scalebox{0.8}{\includegraphics{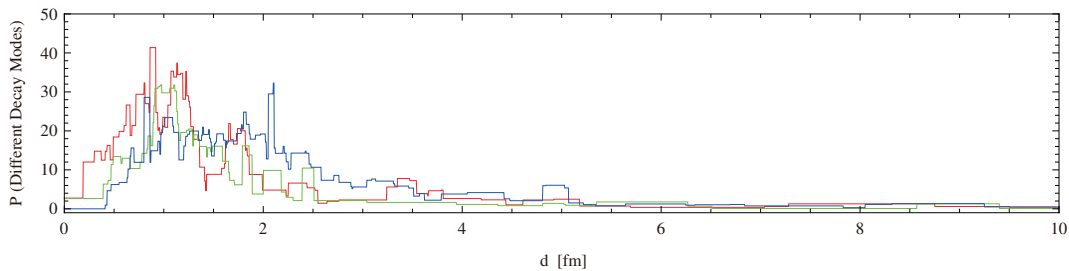}}
\caption{The probability function $P(d)$ for different decay modes. The red, green and blue curves are for $S$-wave, $P$-wave and the other waves, respectively.}
\label{fig:partialwave}
\end{center}
\end{figure}

Based on these results, we would like to discuss the structure of hadrons, and ask several questions:
\begin{enumerate}

\item The first question is whether the time for the final states to separate geometrically should be counted in the decay process. We use the unstable nucleus $^{12}$O as an example. From Tab.~\ref{tab:unstablenuclei}, we find that $d_{^{12}{\rm O}\rightarrow ^{11}{\rm N}p} = 5.26$-$6.08$ fm. Assuming the $^{12}$O nucleus is close-packed, the distance that the outermost proton needs to run away from the rest $^{11}$N nucleus is similar to the proton radius $r_p = 0.84$ fm (or its diameter $2 r_p = 1.68$ fm)~\cite{Pohl:2010zza}. Therefore, in the case of $^{12}$O, its final states, the $^{11}$N nucleus and the proton, have enough time to separate geometrically during its decay process. The same situation happens to other unstable nuclei, and so the answer to the first question can be yes for the case of unstable nuclei.

    Masses of heavy hadrons are of the same order of magnitude as masses of light nuclei, and so we might think this answer is the same for the case of heavy hadrons. However, the distance $d_X$ is less than one femtometer for many hadrons including heavy ones, which number is of the same order of magnitude as the hadronic radii. Moreover, in order to obtain these results we have assumed that the mass difference is totally and immediately transferred into kinetic energies, and so the actual distance that the two final states travel away from each other in the half-life of the initial unstable hadron can be even smaller. In such cases, if we assume that the initial hadron is spherical in the beginning and the two final hadrons are both spherical in the end, the two final states may not separate geometrically even after the whole decay process. Therefore, if the answer to the first question is yes, it seems that many hadrons may have a molecular structure, a non-spherical shape like oval, or even a separated two-body structure like halo nuclei.

\item The second question is whether these short-lived hadrons have already started to decay before they are well formed. This question is related to the first question. If the answers to both questions are yes, the distance $d_X$ can be as small as one femtometer no matter what shapes the initial and final hadrons have. In such cases the initial hadron are not well produced and the two final states behave like being connected by some ``springs'' but they quickly disconnect with each other. However, if the total distance travelled at most, $d_X$ (or $\sim 2 d_X$ starting from their production), is too small, they would become difficult to be observed in experiments.

    Now we try to answer the question raised in the title: how fast can an unstable particle decay into two final states and be observed? From Tab.~\ref{tab:lightmesons} to Tab.~\ref{tab:charmedbottombaryons}, we find that the distance $d_X$ is larger than 0.3 fm for all the hadrons having a lower bound on $d_X$, except $d_{\pi_2(1880) \rightarrow a_2(1320) \eta} = 0.19-0.46$ fm and $d_{K_0^*(800) \rightarrow K \pi} = 0.19-0.37$ fm. However, in these two cases, the theoretical error bars from the uncertainty principle are both large: $\Delta d_{\pi_2(1880)} = 0.49$ fm and $\Delta d_{K_0^*(800)} = 0.76$ fm. Consequently, we use this restriction $d_X = 0.3$ fm to plot several curves in Figs.~\ref{fig:limitation}, where the two final states are set to be $\pi$-$\pi$, $K$-$\pi$, $D$-$\pi$, $D^0$-$D^{*0}$, $N$-$\pi$ and $N$-$K$. The unstable hadrons which can be observed in experiments should be inside the regions $d_X > 0.3$ fm.

    Since this restriction is not so strict, we also plot the curves obtained by fixing $d_X = r_p = 0.84$ fm. Revelent hadrons are shown in these figures for comparisons. The unstable hadrons inside (or not far from) the regions $d_X > r_p = 0.84$ fm can be more easily observed. Particularly, we show $X(3872)$ in the middle-right figure, where $m_{D^0} = 1864.86$ MeV, $m_{D^{*0}} = 2006.99$ MeV, $m_{X(3872)} = 3871.68 \pm 0.17$ MeV and $\Gamma_{X(3872)} < 1.2$ MeV~\cite{Beringer:1900zz}. It is coincident with the beginnings of the curves $d_{X \rightarrow D D^*} = 0.3$ fm and $d_{X \rightarrow D D^*} = 0.84$ fm. From these figures, we quickly notice that these restrictions are more strict on heavy hadrons than on light hadrons.

\item The third question is whether there are still hadrons which have large decay widths but can not be observed due to the small $d_X$. To answer this question, we draw the probability function $P(d) = \sum_X P_X(d)$ at the top and in the middle of Figs.~\ref{fig:totalhadron}, where $P_X(d)$ is defined for each light hadron and each heavy hadron, respectively. For both cases probability functions have maxima when $d_X$ is around one femtometer.
    We also draw the probability function $P(d)$ for different decay modes in Fig.~\ref{fig:partialwave}. We find no significant differences among hadrons decaying through $S$-wave, $P$-wave and the other waves, although it seems that $S$-wave decay patterns should lead to larger decay widths and so smaller $d_X$. These two comparisons suggest that the answer to the third question is yes. Consequently, if we take a look at Fig.~\ref{fig:heavybaryon} again, it suggests that there may be some heavy baryons having decay widths around 100-200 MeV and waiting to be observed.

\end{enumerate}

\section*{Acknowledgments}

This work is supported by the National Natural Science Foundation of China under Grant No. 11205011, and the Fundamental Research Funds for the Central Universities.

\end{document}